\documentclass[fleqn,usenatbib]{mnras}

\usepackage{newtxtext,newtxmath}

\usepackage[T1]{fontenc}

\DeclareRobustCommand{\VAN}[3]{#2}
\let\VANthebibliography\thebibliography
\def\thebibliography{\DeclareRobustCommand{\VAN}[3]{##3}\VANthebibliography}

\usepackage{graphicx}	
\usepackage{amsmath}	
\usepackage{ulem}

\newcommand{\dd}{\mathrm d}
\renewcommand{\d}{\partial}
\renewcommand{\i}{\mathrm i}
 \renewcommand{\(}{\left(}
 \renewcommand{\)}{\right)}
 \renewcommand{\[}{\left[}
  \renewcommand{\]}{\right]}
\renewcommand{\v}[1]{\boldsymbol{#1}}

\usepackage[dvipsnames]{xcolor}

\newcommand{\eVqcm}{\,\mathrm{eV}\,\mathrm{cm}^{-3}}
\newcommand{\ergs}{\,\mathrm{erg}\,\mathrm{s}^{-1}}
\newcommand{\pqcm}{\,\mathrm{cm}^{-3}}

\title[Inefficient particle re-acceleration in massive stellar clusters]{On the inefficiency of particle re-acceleration mechanisms in the cores of massive stellar clusters}

\author[T. Vieu et al.]{
T. Vieu,$^{1}$\thanks{E-mail: thibault@mpi-hd.mpg.de}
L. Härer,$^{1}$
B. Reville,$^{1}$
\\
$^{1}$Max-Planck-Institut f\"ur Kernphysik, Saupfercheckweg 1, D-69117 Heidelberg, Germany
}

\date{Accepted XXX. Received YYY; in original form ZZZ}

\pubyear{2024}

\begin{document}
\label{firstpage}
\pagerange{\pageref{firstpage}--\pageref{lastpage}}
\maketitle

\begin{abstract}
We consider scenarios for non-thermal particle acceleration and re-acceleration in the central cores of compact massive star clusters, aided by insights from high resolution hydrodynamic simulations. We show that i) particles are unlikely to interact with many shocks during their lifetimes in the core; ii) colliding flows do not produce hard spectra; iii) turbulent re-acceleration in the core is suppressed. Inefficient re-acceleration mechanisms are not expected to produce hard components nor to increase the maximum energy within the cores of massive star clusters. Models in which the observed ultra-high energy gamma rays originate in the core of massive stellar clusters are thus disfavoured. 
\end{abstract}




\section{Introduction}
Considerable attention has recently been drawn toward massive star-clusters as efficient particle accelerators and sources of non-thermal radiation, extending up to the ultra-high energy range (> $100$ TeV gamma-ray energies). Although observations in gamma rays from such sources have become increasingly abundant \citep[e.g.,][]{aharonian2019, yang2017, yang2020, gammaCygnus_HAWC2021, cao2021, HESSWd12022}, the field lacks a comprehensive understanding of these regions and in-depth theoretical modelling. In particular, a detailed model that provides reliable predictions adjustable to observational data is still missing.

The concept that massive star clusters could serve as sources of cosmic rays and non-thermal radiation is not new. Over 40 years ago, researchers explored the idea that particles could be accelerated by the stellar winds of powerful early-type stars \citep{montmerle1979, cesarsky1983}, considering it as an alternative to supernova remnant shocks. Not only are massive star clusters capable of producing non-thermal particles, but they may also offer explanations for various abundance anomalies observed in the galactic cosmic-ray spectrum near Earth \citep{gupta2020,tatischeff2021}. Moreover, they could potentially contribute as primary sources in the supra-PeV energy range \citep{vieu2023}.

{The core of a massive star cluster is commonly defined as the region containing half of the stellar mass \citep{portegies2010}. If the cluster is compact enough (typically if the half-mass radius does not exceed a few~pc), wind-wind interactions in the core create complex outflows. The region is filled with an ensemble of strong shocks embedded in a turbulent plasma \citep{badmaev2022}. At the individual shocks, a fraction of the wind material can be accelerated via diffusive shock acceleration. These shocks typically extend over a fraction of a parsec, with a velocity jump of the order of 1000~km/s, which for typical conditions would imply a maximum energy for the accelerated particles to lie \textit{a priori} in the TeV range. On the other hand, it is conceivable that, in such an intricate environment, the outflows of individual stars merge, giving rise to collective effects. These might increase the maximum energy and imprint specific signatures to the spectra of particles and, ultimately, radiation. }

Particle acceleration and propagation in the core of a massive star cluster was previously explored by \citet{bykov1992b}, building on earlier work on particle transport in strong hydrodynamic turbulence \citep{bykov1990}. These models are purely stochastic: the cluster core being described as an ensemble of random shocks on top of a turbulent plasma. A complementary model was proposed by \citet{klepach2000}, considering an ensemble of well-defined stellar wind cavities. Generically, these works show that particle acceleration in the cores of massive star clusters can be not only efficient, but also that collective effects tend, via \textit{re-acceleration} either at shocks or in the turbulence, to harden the particle distribution. On the other hand, including additional physics in the model, such as particle feedback \citep[see][]{bykov2001} or refined geometries \citep[see][]{klepach2000}, affects these conclusions.

The situation is similar in the superbubble cavity inflated a large distance around the cluster core. Using a test-particle model including stochastic and shock re-acceleration, \citet{ferrand2010} showed that a hard ($\sim p^{-3}$) spectrum could develop in this environment. This model was recently revisited in \citet{vieu2022}, where it was shown that the back-reaction of particles onto the turbulence and the shocks reduces the claimed hardening. An important finding was that in superbubble environments characterised by a turbulence level sufficient to confine the particles, the energy density of the particles is expected to be comparable to the energy density of the turbulence and their pressure comparable to that of the shocks. To ensure energy conservation in the system, it is necessary to account for the back-reaction of the particles. This is in particular relevant to cluster cores: test-particle calculations do not provide reliable predictions. In fact, typical spectra from stellar clusters or superbubbles can be steep, already at the stage of first acceleration \citep{morlino2021}. Spectra slightly steeper than $p^{-4}$ are also consistent with the observational data, unlike the hard spectra predicted by those models accounting for collective processes.

Finally, collective acceleration has been claimed to occur in the collision between two shocks. While \citet{bykov2013} argue that a hard spectrum ($\sim p^{-3}$) can develop at the higher end of the spectrum, as discussed in \citet{vieu2020}, this requires that the shocks are both stationary and sufficiently close to each other \citep[see also][]{axford1990}.

{From the aforementioned works, we can identify three re-acceleration mechanisms which can \textit{in principle} impact the distribution of particles that are potentially injected by the ensemble of strong stellar wind termination shocks in the core:}
\begin{enumerate}
    \item \textit{shock re-acceleration}: particles accelerated around a strong shock (e.g. a stellar wind termination shock) reach another strong shock upon diffusing in the turbulent medium. Interactions with multiple shocks within the core \textit{could} progressively harden the particle distribution function.
    \item \textit{colliding shocks}: two nearby powerful stars produce a pair of colliding wind shocks. Particles trapped in the colliding outflows \textit{possibly} develop a hard spectrum as they cannot escape the system.
    \item \textit{stochastic re-acceleration}: particles diffusing in the turbulent plasma scatter on magnetic and hydrodynamic modes, both of which can lead to a second order re-acceleration process which \textit{might} be efficient enough to impart hard signatures at low to intermediate energies.
\end{enumerate}

It is unclear how effective any of these scenarios are in practice, or if they operate at all. In particular, since particles accelerated in the core will be advected over large scales, they are expected to significantly contribute to the diffuse non-thermal emission. However, predictions of hard spectra are in tension with observations, which typically measure a flux scaling as $E^{-2.2\,... -2.6}$ from GeV to TeV.

The cores of compact stellar clusters are complex regions, that remain poorly understood. The goal of the present paper is to investigate the key high-energy non-thermal processes and determine what spectral signatures may be expected. We employ detailed multi-dimensional hydrodynamic simulations (Section~\ref{sec:SimulationsandModel}) to improve our understanding of the flow configuration, which governs particle acceleration and transport in the core. We then examine the three scenarios described above in light of the simulation results. In Section~\ref{sec:MultipleshocksInteractions}, we develop a model for re-acceleration around individual wind termination shocks and we show that it is generically disfavoured that particles efficiently diffuse from the bulk flow back to the strong shocks, making this re-acceleration process inefficient. In Section~\ref{sec:CollidingWinds} we revisit particle acceleration in colliding outflows, accounting for particle escape along the collision plane. We show that a hard component is not expected to develop. In Section~\ref{sec:StochasticReacc}, we demonstrate that feedback from the accelerated particles suppresses stochastic re-acceleration in the turbulent bulk flows. Our conclusions are summarised in Section~\ref{sec:conclusions}.


\section{Modelling massive star cluster cores}\label{sec:SimulationsandModel}
\subsection{Insight from hydrodynamic simulations}
\begin{figure*}
          \centering
              \includegraphics[width=\linewidth]{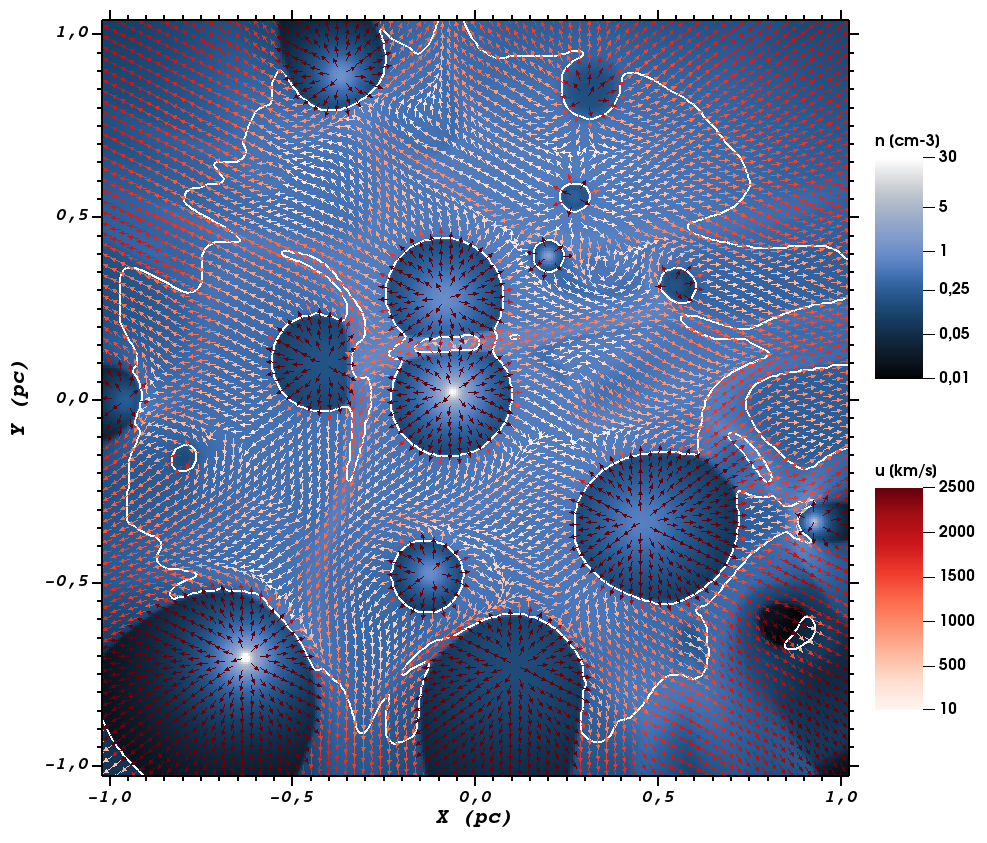}
   \caption{HD simulation of the core of a massive star cluster containing 500 massive stars (93 simulated) within a sphere of radius $1$~pc. Overview of the core in the (X-Y) plane at $t=250$~kyr. The white outlines show the transition between subsonic and supersonic regions.}
   \label{fig:simulation_rho}
 \end{figure*}

\begin{figure}
          \centering
              \includegraphics[width=\linewidth]{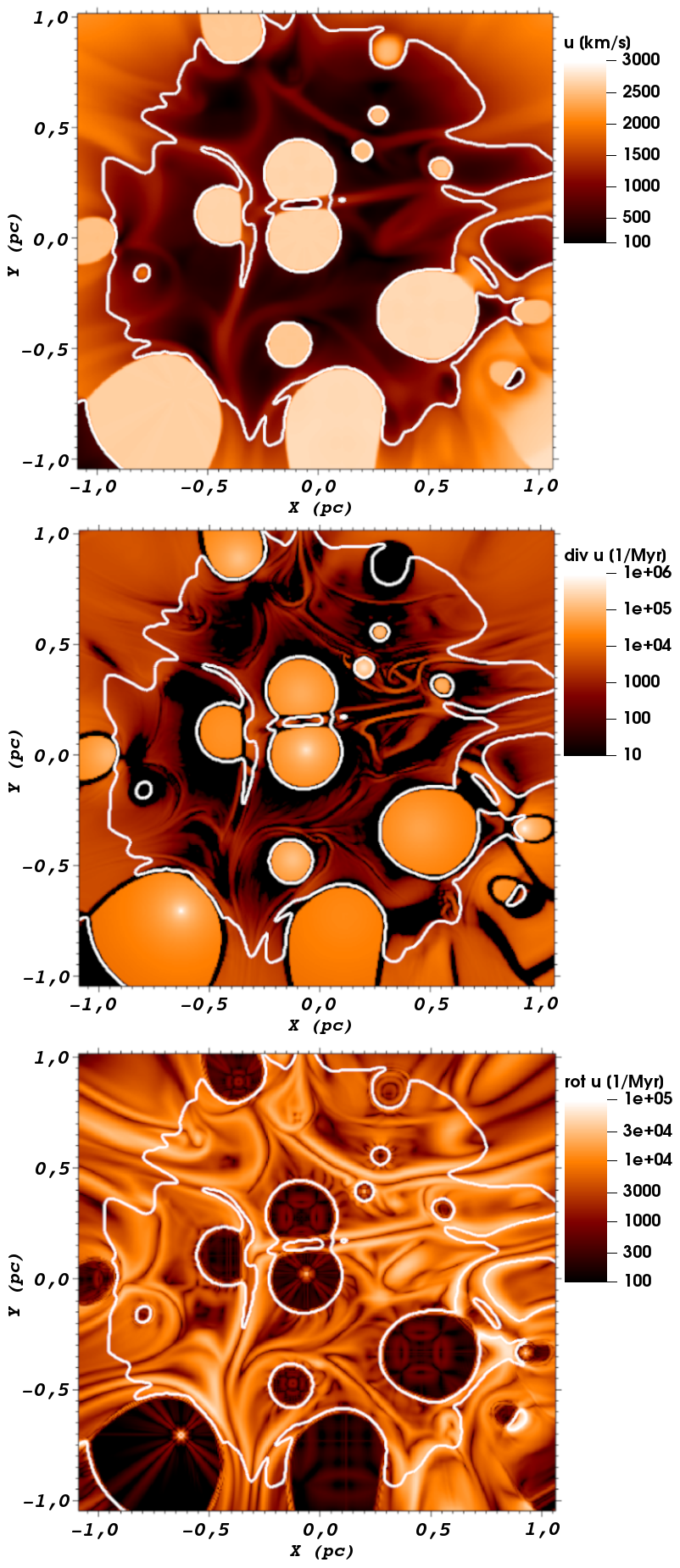}
   \caption{Velocity map from the simulation: norm (top), divergence (middle), norm of vorticity (bottom). The white outline shows the transition between subsonic and supersonic material.}
   \label{fig:velocitydivmap}
 \end{figure}

\begin{figure*}
          \centering
              \includegraphics[width=\linewidth]{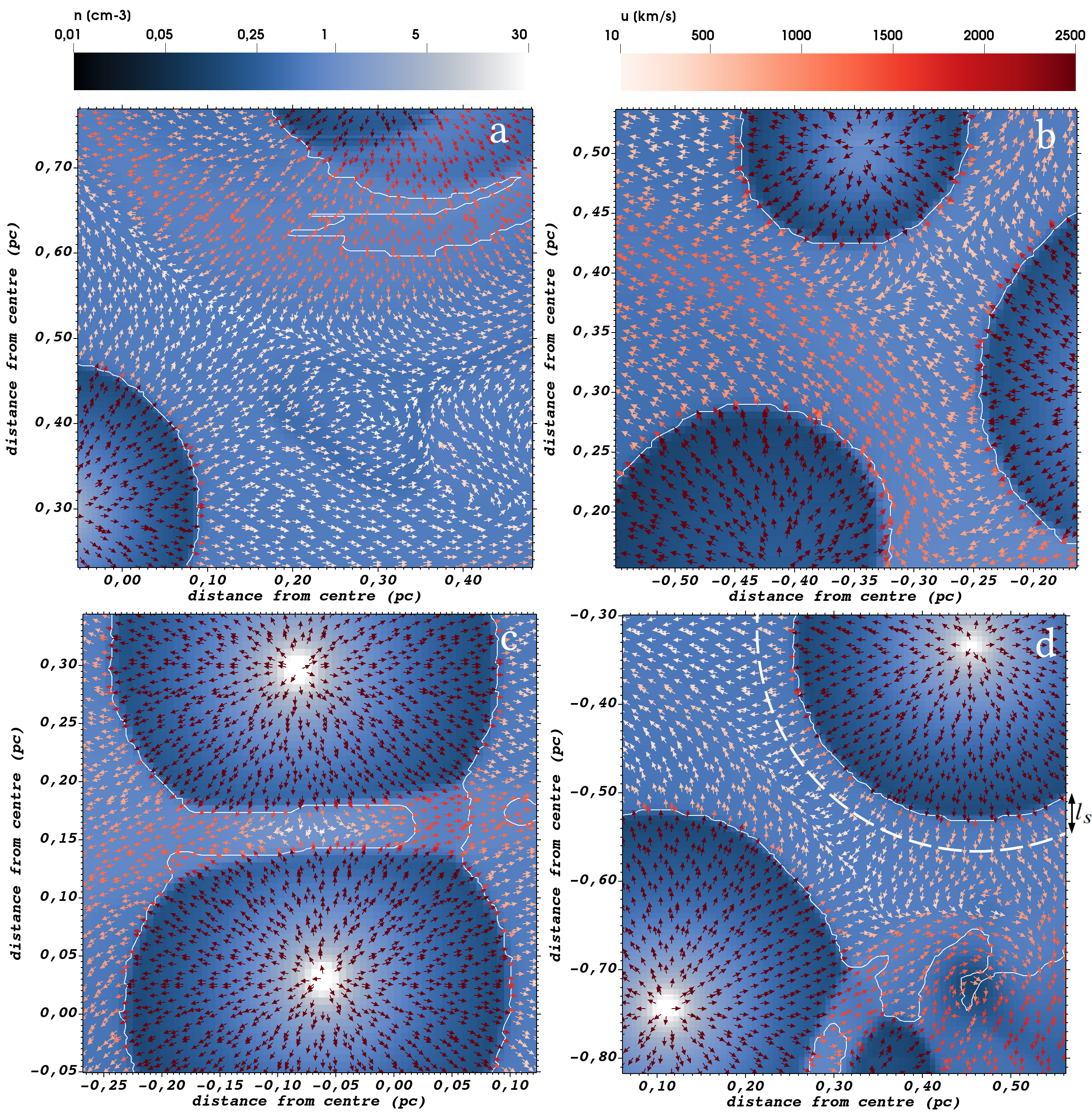}
   \caption{Focus on noticeable features in a simulation slice. a) Mixing of the bulk flows. b) Interaction between three stellar winds. c) Zoom on the system of colliding winds near the origin in a plane crossing the position of both stars. d) Zoom on another system of colliding winds in the bottom right, again in the plane crossing the position of both stars. The white dashed outline in Panel (d) shows the indicative extent of the transition layer $l_s$ defined in Section~\ref{sec:MultipleshocksInteractions}. }
   \label{fig:simulation_zoomed}
 \end{figure*}

\subsubsection{Simulation setup}
We present a high-resolution 3D hydrodynamic (HD) simulation of a cluster core with the publicly available code PLUTO \citep{PLUTO2007}. A stellar population of 500 massive main-sequence stars from 8 to 100~M$_\odot$ was randomly generated by sampling a $M^{-2.35}$ initial mass function \citep{salpeter1955,kroupa2002}. Individual stars were assigned random positions within a sphere of radius 1~pc, following a homogeneous distribution (not considering possible mass segregation). Stellar winds are then setup within spheres of radius 0.05~pc (8 cells). Only stars with a mass larger than 25~M$_\odot$ were included in the simulation, as stars with lower masses were found not powerful enough to expand a wind beyond the injection region, the size of the latter being constrained by the resolution of the simulation, $\Delta x = 5.9\times 10^{-3}$~pc. The resolution was chosen in order to ensure that there are at least 32 cells between two stars, which is necessary to avoid artificial wind-wind interactions. The high-resolution grid is only applied in a cube of edge-length $2 \times 1.184$~pc, beyond which the grid is progressively stretched up to the boundary of the domain. In this way, we cover a large cubic volume which extends from -20~pc to 20~pc in each of the three Cartesian directions, with 512$^3$ cells. Although we are only interested in the cluster core, it is found to be necessary that the superbubble blown by the cluster remains inside the domain, otherwise the pressure beyond the cluster core would be underestimated and the results become unreliable. We adopt an initial external density of 50~cm$^{-3}$, to approximate the giant molecular cloud inside which the cluster was born. The superbubble shell reaches the boundary of the domain after about 300~kyr. By that time, the cluster core has reached a quasi-stationary state and is not expected to evolve until stars start to transition to their late stages of evolution, which we do not consider here.

We used simple prescriptions for the stellar wind velocities and mass loss rates from fits of stellar evolution models \citep[][for non-rotating, single stars]{seo2018}:
\begin{align}
    &\log_{10} \( \frac{\dot{M}}{{ M_\odot/yr}}\)
    = -3.38 \( \log_{10} \frac{M}{M_\odot} \)^2 + 14.59 \log_{10} \frac{M}{M_\odot} - 20.84 \, ,
    \label{stellarparametersMdot}
    \\
    &\log_{10} \( \frac{V_w}{{\rm km/s}}\) = 0.08 \log_{10} \frac{M}{M_\odot} +3.28 \, .
    \label{stellarparametersVw}
\end{align}
The total power of the simulated cluster is $P_* = 1.2 \times 10^{38}$~erg/s, for a total mass loss rate $\dot{M} = 1.1 \times 10^{-4}$~M$_\odot$/yr. The velocity and density profiles imposed within the injection cells around the $i$-th star are:
\begin{align}
    \v{u} &= u_{w,i} \v{e_r} \, , \\
    \rho &= \frac{\dot{M}_i}{4 \pi r^2 u_{w,i}} \, ,
\end{align}
where $u_{w,i}$ and $\dot{M}_i$ are obtained as functions of the zero-age-main-sequence mass using Eqs.\ref{stellarparametersMdot},~\ref{stellarparametersVw},
while the pressure is prescribed assuming an isothermal wind: $p_{\rm gas} = \rho c_s^2$, with $c_s = 20$~km/s.

To minimise the computational overhead in our high resolution simulations, we neglect thermal conduction and cooling.

\subsubsection{Simulation results}
We show the results of the simulation at time $t=250$~kyr in a representative slice through the domain.
Fig.~\ref{fig:simulation_rho} shows the density map over the entire core, while Fig.~\ref{fig:velocitydivmap} displays the properties of the velocity field. The core can be divided into three regions: the stellar wind cavities, the subsonic bulk and the supersonic bulk. First, the ensemble of stellar wind cavities are low density regions, each of them typically extending over a few tenths of a pc and characterised by a low temperature radial, supersonic outflow with a typical velocity of a few thousand km/s. These regions are delimited by the individual wind termination shocks (WTS), which remain approximately spherical. Then, downstream of the stellar wind cavities, the outflows combine to form a hot, turbulent, plasma with a typical mean-square velocity of a few hundred km/s. There the plasma is subsonic i.e. nearly incompressible, and turbulent, as can be seen in the middle and bottom panels of Fig.~\ref{fig:velocitydivmap}. Finally, at the edge of the core, the bulk flow becomes supersonic again and cools to give rise to the ``cluster wind'' theoretically expected \citep{weaver1977}. 

Fig.~\ref{fig:simulation_zoomed} shows several noticeable features in a different plane. Panel (a) shows how the downstream flows combine to form a turbulent bulk (notice the large-scale eddy in the lower right). Similarly, the top right panel shows how three strong winds impact on the collective bulk flow in a plane shifted with respect to the position of the stars. Note that, in the downstream vicinity of the shocks, the flows typically remain radial over a length $l_s \approx 0.01-0.1$~pc (see e.g. the white dashed outline in Panel (d)). In anticipation of Section~\ref{sec:MultipleshocksInteractions}, this implies that particles will only interact with multiple shocks if their diffusion length is larger than $l_s$, i.e. multiple shock interactions should only be expected for the highest energy particles.

The lower left panel of Fig.~\ref{fig:simulation_zoomed} shows a stationary collision between two stellar winds. Both outflows decelerate, creating a stagnation point, from which matter is evacuated in the direction orthogonal to the collision axis. The velocity profile along the collision axis scales approximately linearly with the distance from one shock to the other, as does the outflow escaping in the orthogonal plane. Anticipating Section~\ref{sec:CollidingWinds}, the presence of a stagnation point implies that 
advection in the downstream will sweep particles out of the system before they are energetic enough to interact with both shocks. The latter would require the particle's gyroradius to exceed the length scale of the downstream flow. 

Finally, the lower right panel shows a more complicated collision: the stars are not as close and the downstream region is influenced by the bulk flows. Another strong eddy can be seen at the edge of the core in the lower right corner.

\subsubsection{Modelling the hydrodynamic properties of the core}
\begin{figure}
          \centering
              \includegraphics[width=\linewidth]{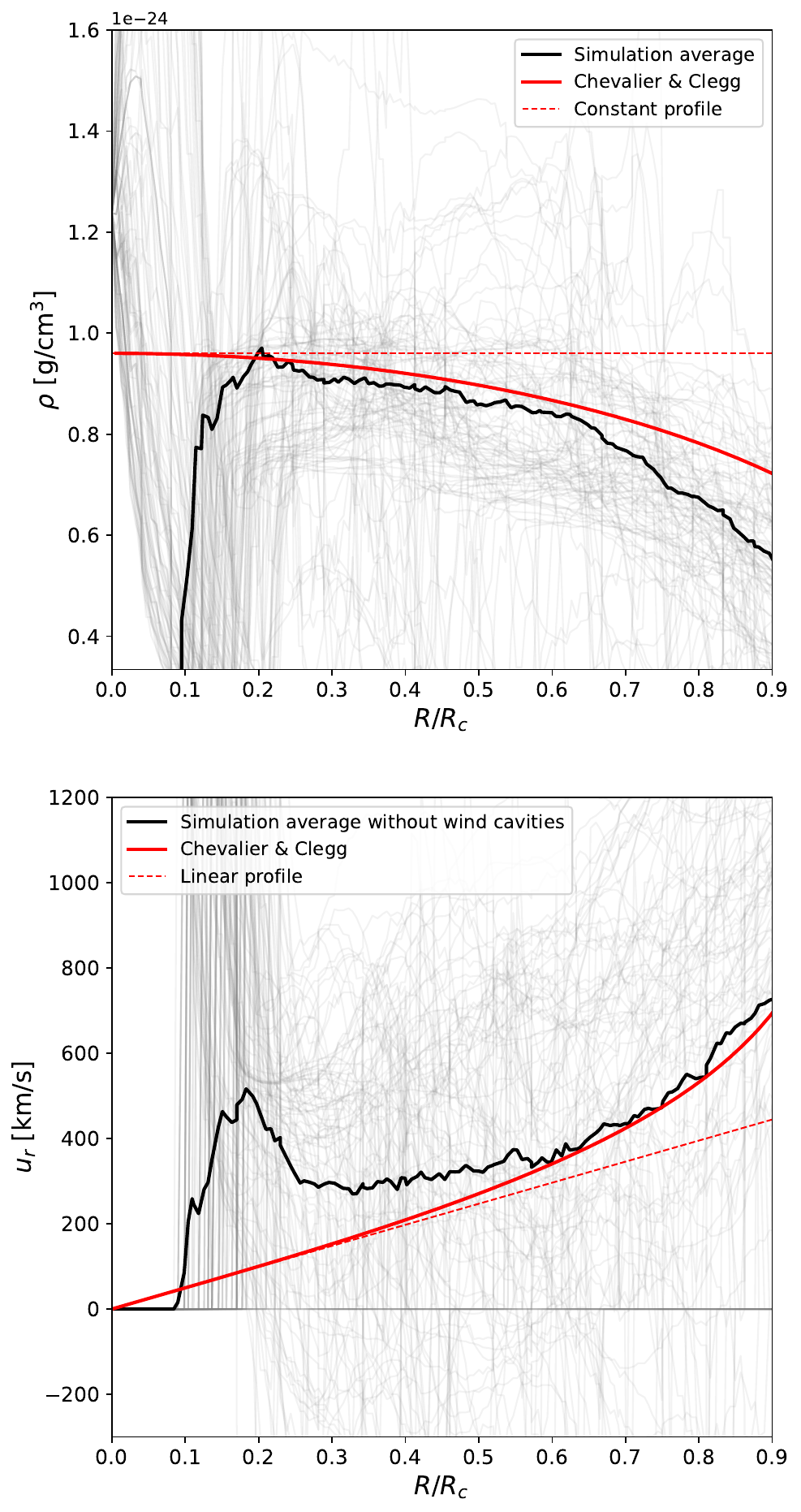}
   \caption{Radial profiles of density (up) and radial velocity (down) from our simulation. The solid black line shows the average over 500 sampled line-outs (shown by the light grey lines). Each point for the radial velocity is obtained by averaging over a cube of sidelength 0.05~pc in order to get rid of the turbulent motions. The wind cavities are partially removed based on a threshold on the divergence, vorticity and temperature of the plasma.
   }
   \label{fig:radialprofiles}
 \end{figure}

We model the core as a symmetric sphere with radius $R_c$ over which a number $N_*$ of powerful early-type stars are distributed homogeneously. The stellar winds are shocked at some typical distance $R_s$ from the stars and the outflows merge in the downstream region. The core therefore contains an ensemble of strong shocks and random outflows. As in the numerical model, we refer to the region upstream of the shocks as the ``wind cavities'', and the common downstream as the ``bulk''. There is a transfer of material from the wind cavities to the bulk, but not from the bulk to the cavities.

The bulk velocity averaged over solid angle is denoted $\bar{u}$ and scales approximately as $\bar{u} \propto r$, while the average density is approximately constant over the region of energy injection \citep{chevalier1985,canto2000}:
\begin{align}\label{rhouchevalier}
    \bar{\rho}_c &= 
     \frac{1}{2 \pi \epsilon} \frac{P_*}{u_w^3 R_c^2} \, ,
    \\
    \bar{u} &= u_w \epsilon \frac{r}{R_c} \, ,
\end{align}
where $\epsilon \approx 0.19$ for $r \ll R_c$. Fig.~\ref{fig:radialprofiles} shows the radial profiles of the density and velocity obtained from the simulation, compared to the solution of \citet{chevalier1985} and the above approximation. Note that the latter is not obtained using the cluster wind velocity, $\sqrt{2 P_*/\dot{M}} \approx 1700$~km/s, but using the average velocity $u_w \approx 2600$~km/s. Overall we find reasonable agreement between theory and simulation. A perfect match is not expected as the theory of \citet{chevalier1985} is based on the hypothesis of continuous energy deposition over the core, while in our simulation there is a strong inhomogeneity which does not average out. This is most evident near the origin due to a concentration of strong colliding outflows. Nevertheless, we adopt the prescription given by Eq.~\ref{rhouchevalier} in the present work as it provides a reasonable estimate of the average density and bulk velocity.
Eventually, the velocity averaged over the entire core is obtained as $\bar{u}_c = 3/4 u_w \epsilon \approx 0.14 u_w$.

\begin{figure}
          \centering
              \includegraphics[width=\linewidth]{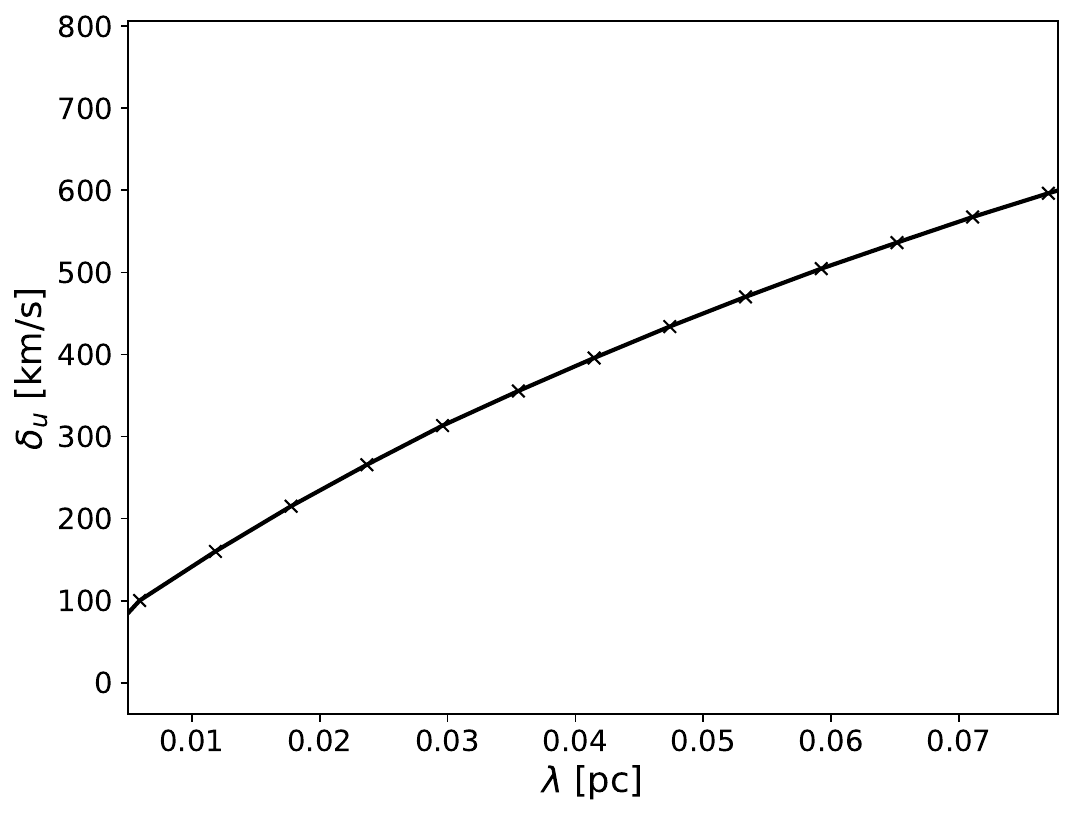}
   \caption{Turbulent velocity in the bulk calculated from the second order structure function (Eq.~\ref{S2function}).}
   \label{fig:radialprofilesdeltau}
 \end{figure}

The colliding outflows transport stellar magnetic fields and generate turbulence in the bulk, from an injection scale $L_T$ of the order of the size of the large-scale eddies: $L_T \approx 0.1$~pc. It can be seen in the bottom panel of Fig.~\ref{fig:velocitydivmap} that the vorticity is high in the bulk, typically in the range $10^3 - 10^5$~Myr$^{-1}$. Interpreting this value as the eddy turnover time close to the injection scale, one concludes that the turbulent velocity $\delta u$ should be in the range $100 - 1000$~km/s, i.e. a sizeable fraction of the average bulk velocity. In order to probe the turbulence more quantitatively, we calculate the second order structure function \citep[e.g.][]{Frisch1995}:
\begin{equation}\label{S2function}
    \mathcal{S}_2(\lambda) = \langle | \v{u}(\v{r}) -  \v{u}(\v{r} + \v{\lambda}) |^2 \rangle
= \delta u ^2 (\lambda) \, ,
\end{equation}
where $\lambda = | \v{\lambda} |$. For each position $\v{r}$ the average of $| \v{u}(\v{r}) -  \v{u}(\v{r} + \v{\lambda}) |^2$ is performed over all points located at a distance $\lambda$ from $\v{r}$, which gives the turbulent velocity field, eventually averaged over the volume of the bulk.
This probes the average turbulence strength in the bulk at a given scale $\lambda$, whose square root can be physically interpreted as the turbulent velocity $\delta u$. As shown in Fig.~\ref{fig:radialprofilesdeltau}, $\delta u$ lies in the range $300 - 500$~km/s at scales comparable to the injection scale.

This can be compared to the usual phenomenological estimate based on the assumption that a fraction $\eta_T$ of the wind kinetic energy downstream of the shocks is transferred into MHD turbulence and that magnetic and hydrodynamic modes are in equipartition:
\begin{equation}\label{deltauequipartition}
    \delta u = v_A = \(\eta_T/2\)^{1/2} u_w/4 \, .
\end{equation}
In order to obtain $\delta u \approx 300 - 500$~km/s one needs $\eta_T \gtrsim 50 \%$, i.e. highly efficient generation of turbulence, which is expected given the numerous wind-wind interactions occurring in the compact and powerful cluster that we simulate.

Assuming a mass density of $10^{-24}$~g/cm$^{3}$ in the cluster (top panel of Fig.~\ref{fig:radialprofiles}), we can estimate an average magnetic field $B_c \sim 100$~\textmu G, in agreement with the results of previous MHD simulations \citep{badmaev2022,haererICRC23}.

While the spatial resolution in our simulation is not sufficient to probe the inertial range of the turbulence, it is expected that in reality the turbulent kinetic-energy cascades over an ensemble of scales. In the inertial range, the energy distribution can be modelled with a power-law spectrum:
\begin{equation}
    \delta u^2 = v_A^2 = 2 \pi \int_{k_T}^\infty \dd k k^2 S(k)
    \, , \quad 
    S(k) \propto k^{-\nu} \, ,
\end{equation}
where $k_T=2\pi/L_T$ is the wavenumber associated to the injection scale introduced earlier.

The parameters discussed above describe the turbulent environment in which particles are accelerated and propagate. Overall we find a good agreement between the simulation, calculations and phenomenological arguments.

\subsection{Particle acceleration and transport}
\subsubsection{Particle acceleration}
Diffusive shock acceleration is expected to take place around the strong shocks at the interfaces of the wind cavities of the most powerful stars. A fraction of the upstream wind is, in principle, injected into a non-thermal particle distribution. As a zeroth order approximation, we adopt a standard power-law solution, $f(p) \propto p^{-s}$ for particles with diffusion length much smaller than the size of the shocks. At these energies, the shocks are approximately planar, and particles are confined close enough to the shock to not be affected by the bulk outflows. The possible modulation of the spectrum in the higher energy bands will be tackled in Section~\ref{sec:MultipleshocksInteractions}.

According to diffusive shock acceleration theory, the non-thermal tail develops at most up to the energy where the diffusion length of the particles becomes of the order of the size of the shocks. In the Bohm diffusion limit, this approaches the Hillas limit \citep{hillas1984}. The shocks can be assumed to have a mean curvature $1/R_s$, where $R_s \approx 0.1$~pc is the typical distance from a single star to its wind termination shock. If the diffusion length is equal to $R_s$, the particles which are advected downstream have little chance to diffuse back to the curved shock, which produces an exponential cutoff at the maximum momentum $p_m$ such that $D(p_m) = u_w R_s/\sigma$, where $\sigma \approx 4$ is the compression factor of the shock \cite[e.g.][]{BK88}. Assuming typical shock values $u_w = 2500$~km/s, $R_s = 0.1$~pc, and adopting an optimistic assumption of Bohm diffusion in an average magnetic field $B_c = 100$~\textmu G, we find a maximum momentum $p_m \lesssim 100$~TeV/$c$.

The particle distribution $f(p) \sim p^{-s} e^{-p/p_m}$ which develops around each shock is then advected downstream, in the bulk, where it propagates before ultimately being transported beyond the core of the stellar cluster.

\subsubsection{Transport in the bulk}\label{sec:transportinbulk}
The transport in the bulk, a diffusion length away from the strong shocks, is driven by the turbulence. Particles scatter frequently on non-stationary magnetic fluctuations, which leads to both diffusion in space and diffusion in momentum. The latter is usually referred to as ``stochastic re-acceleration'' and will be discussed in detail in Section~\ref{sec:StochasticReacc}. The transport equation in small-scale turbulence reads:
\begin{equation}\label{TEsmallscaleturbulence}
    \d_t f + \v{u} \cdot \nabla f = \nabla \kappa \nabla f + \frac{\nabla \cdot \v{u}}{3} p \d_p f + \frac{1}{p^2} \d_p p^2 \kappa_p \d_p f \, ,
\end{equation}
where $\v{u} = \v{\bar{u}} + \v{\delta u}$ is the sum of the average bulk velocity and the turbulent velocity, $\kappa \approx \frac{v}{3} L_T \left( \frac{r_L}{L_T} \right)^q$
is the spatial diffusion coefficient on the magnetic waves, with $v$ and $r_L$ respectively the velocity and Larmor radius of the particles, and $\kappa_p \propto p^2 v_A^2/\kappa$ is the momentum diffusion coefficient, which is responsible for particle re-acceleration in the bulk. On average, $\nabla \cdot \v{u} \neq 0$ in the bulk (see the middle panel of Fig.~\ref{fig:velocitydivmap}). This is expected since the large-scale outflow accelerates toward the edge of the core ($\bar{u} \propto r$). Compression and rarefaction waves in the bulk also induce local variations of the turbulent component $\v{\delta u}$, which will lead to stochastic variations of the energy of the particles and eventually a correction to the momentum diffusion coefficient $\kappa_p$. Similarly, the term $\v{\delta u} \cdot \nabla f$ is responsible for random motion of the particles stuck in the eddies, which might bring a correction to the spatial diffusion coefficient $\kappa$. These two processes arise in the so-called ``strong turbulence regime'', $\kappa \ll \delta u L_T$, and was tackled by \citet{bykov1990a} in a two-step procedure. First, Eq.~\ref{TEsmallscaleturbulence} is averaged over the turbulent scales using standard perturbation theory, which provides corrections of the form $D_p = \kappa_p + \delta D_p$, $D = \kappa + \delta D_x$, where $\delta D_x \lesssim \delta u L_T$ and $\delta D_p = p^2 \delta u^2/9 \kappa$ \citep[see also][]{ptuskin1988}. Second, the result is renormalized to account for strong perturbations. Eventually, the transport equation for the distribution function averaged over the turbulence $\bar{f}$ is obtained in the following form:
\begin{multline}\label{TEturbulenceaveraged}
    \d_t \bar{f} + \bar{\v{u}} \cdot \nabla \bar{f} = \nabla D \nabla \bar{f} + \frac{\nabla \cdot \bar{\v{u}}}{3} p \d_p \bar{f} + \frac{1}{p^2} \d_p p^2 D_p \d_p \bar{f} \, .
\end{multline}
The second term on the LHS encodes the advection in the large-scale velocity field which, on average, flows out of the cluster core and provides the main escape channel for low energy particles. The first term on the RHS models spatial diffusion on both the magnetic and hydrodynamic modes of the turbulence, with a total spatial diffusion coefficient $D$. The second term on the RHS accounts for adiabatic losses in the bulk due to the diverging large-scale flow. The last term on the RHS accounts for stochastic re-acceleration of the particles interacting with both magnetic and hydrodynamic modes, with a total momentum diffusion coefficient $D_p$. The renormalised transport coefficients read \citep{bykov1990a}:
\begin{align}
D &= \kappa + \frac{4 \pi}{3} \Re  \iint \dd k \dd \omega k^2 \[ \frac{3 S}{i \omega + D k^2} - \frac{2 D k^2 S}{(i \omega + D k^2)^2} \] \, , \label{chiBykov}
\\
D_p &= \kappa_p + \frac{4 \pi p^2}{9} \Re  \iint \dd k \dd \omega \frac{k^4 S}{i \omega + D k^2}
\, . \label{DpBykov}
\end{align}
In the following we assume a power-law spectrum with Lorentzian time correlations. The timescales associated with spatial diffusion, stochastic re-acceleration and advection are respectively:
\begin{align}
    \tau_D = \frac{R_c^2}{\alpha D}
\, , \quad
    \tau_p = \frac{p^2}{2 D_p}
\, , \quad
    \tau_A = \frac{R_c}{4 \bar{u}_c}
\, ,
\end{align}
where $\alpha \approx 8$ for particles homogeneously injected inside a sphere (this number was obtained by performing a Monte-Carlo simulation of particles uniformly injected in a sphere and diffusing up to the edge).
%
\begin{figure}
          \centering
              \includegraphics[width=\linewidth]{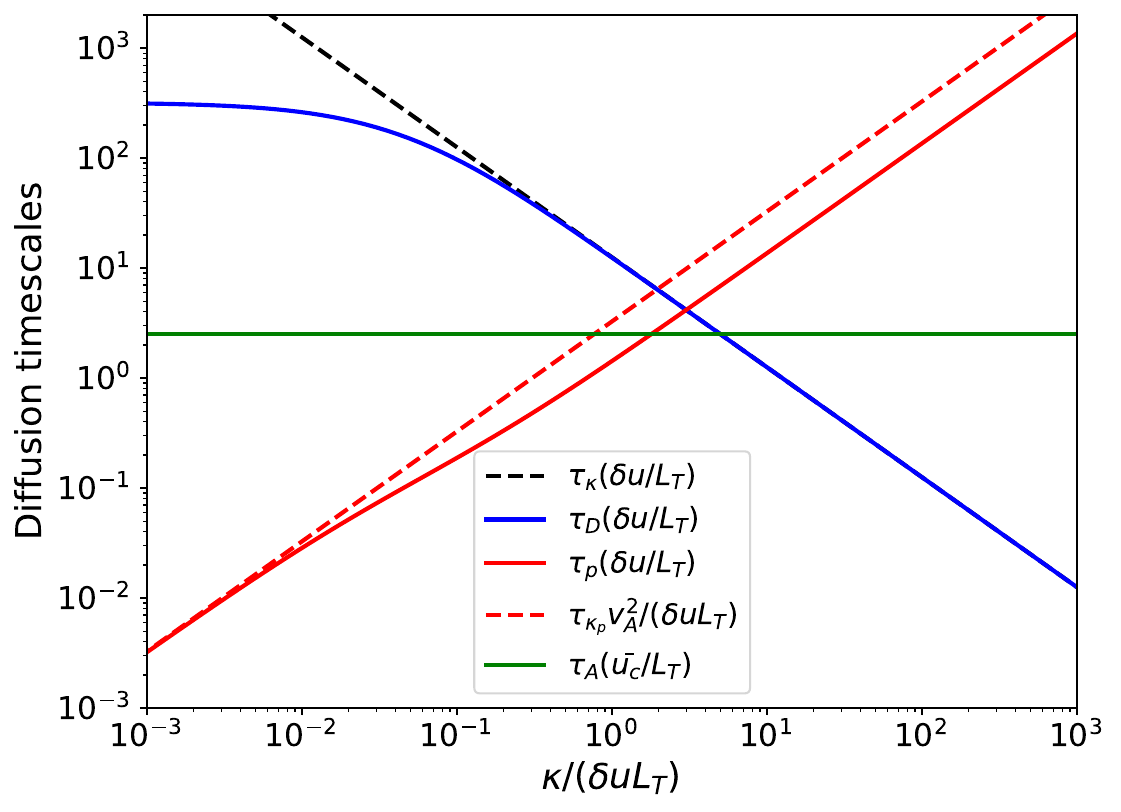}
   \caption{Timescales of diffusion in the turbulence in the case of a Kolmogorov turbulence spectrum, assuming $R_c = 10 L_T$. Dashed-black: spatial diffusion without accounting for the hydrodynamic correction. Solid blue: spatial diffusion including the hydrodynamic modes. Dashed red: momentum diffusion on the magnetic modes. Solid red: momentum diffusion accounting for the hydrodynamic correction. Green: large-scale advection.}
   \label{fig:timescales_core_PLL}
 \end{figure}
These timescales are shown in Fig.~\ref{fig:timescales_core_PLL} in the case of a Kolmogorov spectrum of turbulence (the result is similar in the case of a Kraichnan spectrum). 

Here we wish to highlight a number of points. First, if $v_A \approx \delta u$, i.e. if the magnetic and hydrodynamic modes are in equipartition, accounting for the hydrodynamic modes in the calculation of the momentum diffusion coefficient only changes the latter by a factor of at most two. Only if the magnetic modes are suppressed, $v_A \ll \delta u$, will the correction become more important. We assume in the following that this is not the case, as massive stars typically generate strong magnetic fields and one can expect the equipartition $v_A \approx \delta u$ to hold in the bulk. Second, one can note that for $\kappa \gg \delta u L_T$, the transport is dominated by the diffusion on the magnetic waves. In this case, the particles are too energetic to be affected by the compression and rarefaction waves and $D = \kappa$. On the other hand, in the regime $\kappa \lesssim \delta u L_T$, the transport is dominated by the large-scale advection toward the edge of the cluster and therefore the value of the diffusion coefficient is irrelevant.

In conclusion, in the case of a typical compact cluster core, with $R_c \approx 10 L_T$, $v_A \approx \delta u \sim \bar{u}_c$, one can simply assume $D = \kappa$ and $D_p \approx 2 \kappa_p \propto v_A^2 p^2/\kappa$. One can check that our results are reliable whether or not the transport coefficients are properly calculated or the above approximation is used.

The injection of non-thermal particles in the bulk is modelled by a source term which is added to the transport equation~\ref{TEturbulenceaveraged}:
\begin{equation}
    Q = \sum_{i=1}^{N_*} \delta(\v{r} - \v{r_i}) \phi_s \, ,
\end{equation}
where $\phi_s$ is the flux of particles at the interface with a wind cavity, which contains the particle distribution function advected downstream of the strong shocks, $f(p) \propto p^{-s} e^{-p/p_m}$, as well as possible shock re-acceleration mechanisms which will be discussed in detail in the next section. The stars around which efficient acceleration takes place are located at the positions $\{ \v{r_i} \}$.

We eventually average the transport equation over the volume of the core in order to obtain the following equation for the cluster distribution function $f_c$:
\begin{multline}\label{TEbulkaveraged}
    \d_t f_c + \frac{f_c}{\tau_e} =  \frac{4 \bar{u}_c}{3 R_c} p \d_p f_c + \frac{1}{p^2} \d_p p^2 D_p \d_p f_c + \frac{3 N_*}{4 \pi R_c^3} \phi_s \, ,
\end{multline}
where the escape time $\tau_e$ contains the contribution of advection and diffusion:
\begin{equation}\label{escapetime}
    \tau_e = \(\frac{4 \bar{u}_c}{R_c} + \frac{\alpha D}{ R_c^2} \)^{-1}
    \, .
\end{equation}
As before, $\alpha \approx 8$ for particles homogeneously injected inside a sphere.

\begin{figure}
    \centering
    \includegraphics[width=\linewidth]{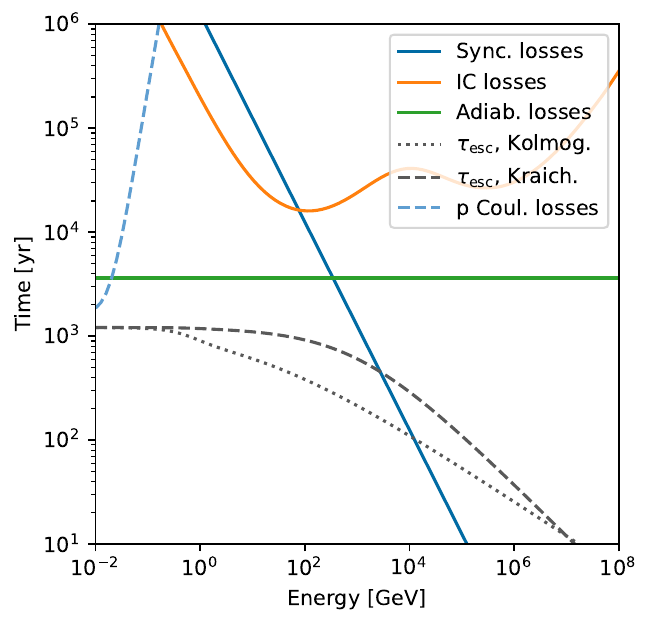}
    \caption{Timescales of escape and losses in the core. The synchrotron, bremsstrahlung, and inverse Compton (IC) losses are computed using the \texttt{GAMERA} library 
    \citep{hahn22}. Coulomb losses are calculated according to \citet{schlickeiser_book}.}
    \label{fig:loss_esc}
\end{figure}

\subsubsection{Losses}
The core of a massive star cluster is filled with powerful radiation fields and strong magnetic fields. Nevertheless, as shown in Fig.~\ref{fig:loss_esc}, the escape time is smaller than the loss time at energies $\lesssim 10$~TeV. Inverse Compton (IC) losses include direct and dust-scattered starlight \citep[][]{Popescu17} at a fiducial galacto-centric distance of 5\,kpc, the Cosmic Microwave Background ($0.25 \eVqcm$) and a thermal cluster photon field with a temperature of $10^4\,$K. 
The total luminosity of the cluster is assumed to be $L_\mathrm{bol} \approx 100 P_*$, which gives an energy density of $U_\mathrm{cl} \sim 1680 (P_*/10^{38}\ergs) (R_\mathrm{c}/1\,\mathrm{pc})^{-2}\eVqcm$. While such a large energy density in principle results in strong losses, the expected short wavelengths of the cluster photons results in Klein-Nishina (KN) suppression of the cooling above $E_{\rm KN} = 75 (T/10^4\,\mathrm{K})^{-1}$\,GeV \cite[e.g.][and Fig.~\ref{fig:loss_esc}]{BG70}. For bremsstrahlung and Coulomb losses, we assumed a typical density $n \sim 0.5\pqcm$. 
The synchrotron losses are calculated assuming an average magnetic field of 100~\textmu G over the core. Synchrotron losses of electrons might be relevant beyond 10~TeV if the magnetic field is large ($B \gtrsim 100$~\textmu G), however this will only affect the  last decade of the particle distribution function and we do not include it in our calculation.

\section{On the inefficiency of particle re-acceleration by multiple shocks}\label{sec:MultipleshocksInteractions}
The goal of this section is to reconsider the scenario in which particles interact with multiple shocks in the cores of massive stellar clusters. The possibility was first raised by \citet{bykov1992b} using a purely stochastic model. The authors showed that hard spectra (asymptotically scaling as $p^{-3}$) could be obtained upon efficient re-acceleration. The idea was then revisited by \citet{klepach2000} considering individual wind cavities with energy-independent diffusion in the downstream region. In this scenario, efficient re-acceleration could only be obtained assuming a large diffusion coefficient. Indeed, if the shocks are not purely stochastic, particles cannot access the upstream wind via advection, as implicitly assumed in \citet{bykov1992b}, but only by diffusion.

In the following we propose an alternative model, assuming the plasma is cold and radial within the wind cavities, and consider the possibility that particles might be advected close to the shocks in large-scale bulk flows.

\subsection{Model}
We consider that the shocks are locally approximately spherical with a curvature of order $1/R_s^2$, where $R_s$ is the radius of a single WTS. 
Immediately downstream of the WTS, the outflow from the nearby star dominates over a transition layer of thickness $l_s \approx 0.1 R_s$, as can be seen in our simulation (Figures~\ref{fig:simulation_rho}, \ref{fig:simulation_zoomed}, see in particular the white dashed outline in Panel (d) of Fig.~\ref{fig:simulation_zoomed}). 
Beyond the transition layer, the bulk flow takes over. It is still expected that the hot, shocked, flow is almost incompressible, however the velocity field is not purely radial anymore. The top right panel of Fig.~\ref{fig:simulation_zoomed} shows such a large-scale flow extending from the shock in the bottom right to the shock at the top. Although the spherical symmetry of the wind is broken beyond the transition layer, particles propagating into this region could still reach the shock via diffusion, unless their diffusion length in the radial component of the flow is smaller than the distance to the shock. In the latter case, particles are screened by the transition layer.

\begin{figure}
          \centering
              \includegraphics[width=0.8\linewidth]{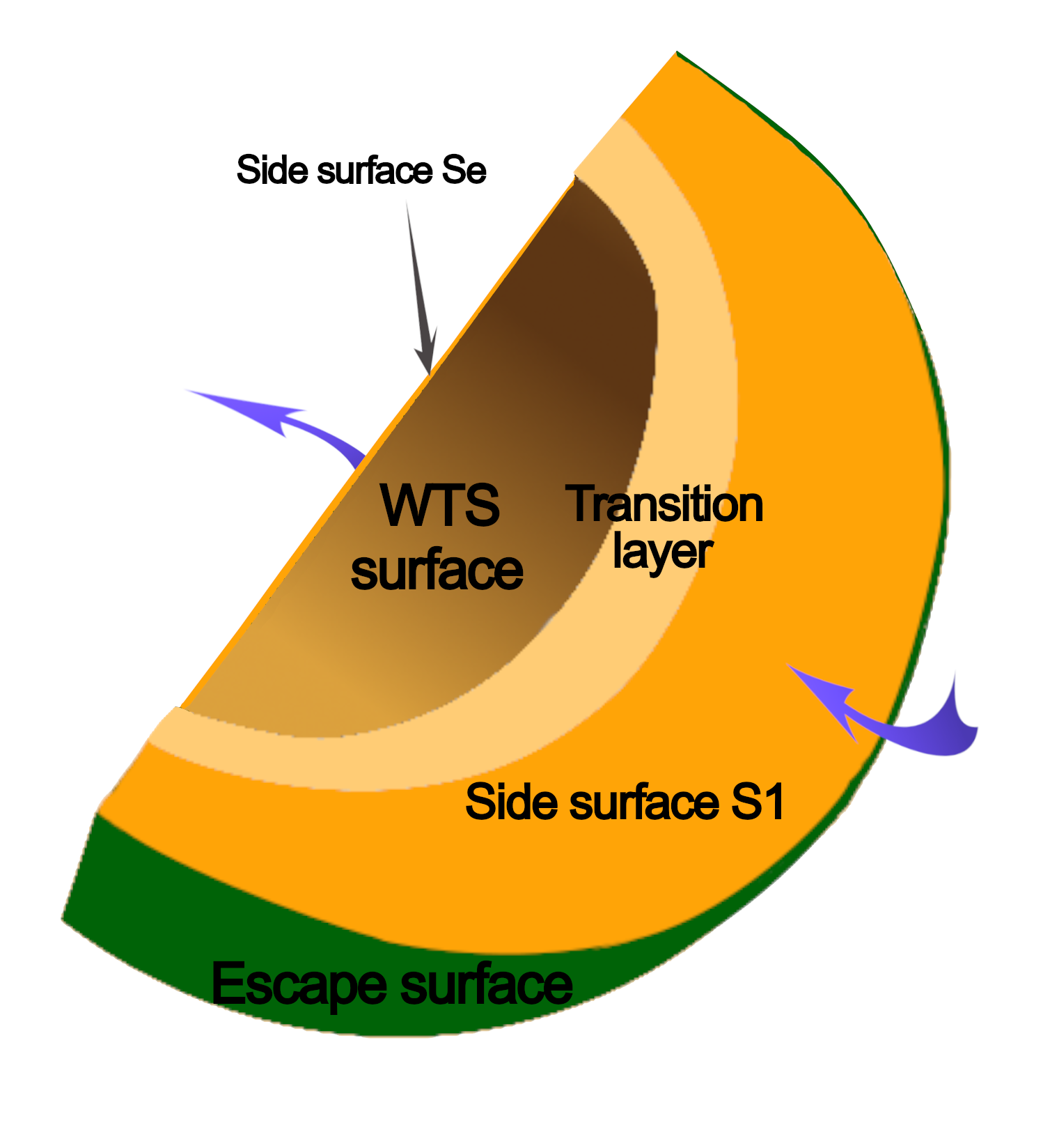}
   \caption{Sketch of the model for shock re-acceleration, showing a slice of the WTS close to an azimuthal bulk flow (purple arrows). Particles advected in this bulk flow enter the region between the WTS surface and the escape surface and can then be re-accelerated. Particles escape the accelerator either by diffusing beyond the escape surface or by being advected in the azimuthal direction.}
   \label{fig:scheme_surfaces}
 \end{figure}

\begin{figure*}
          \centering
              \includegraphics[width=1\linewidth]{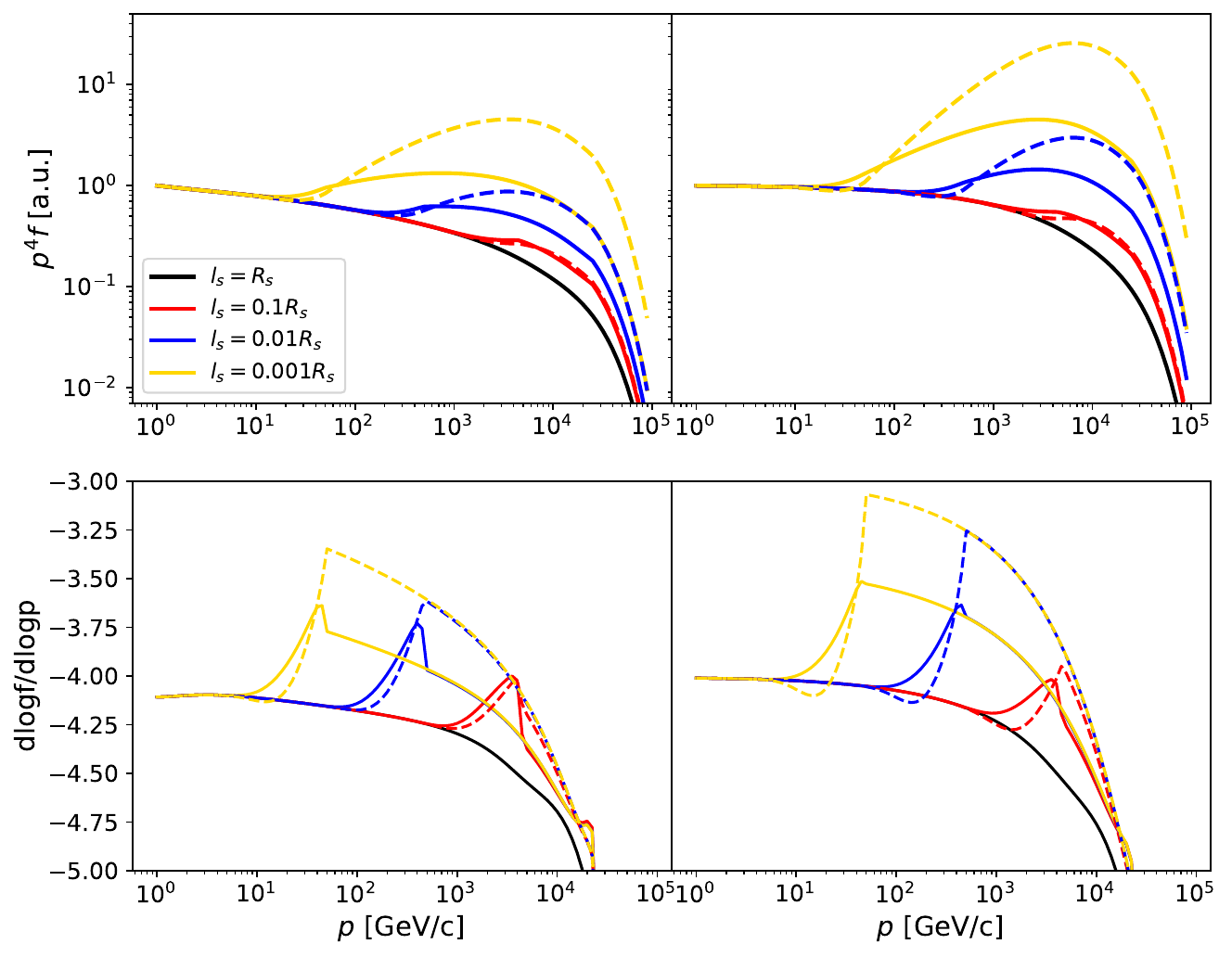}
   \caption{Average spectra obtained via multiple shock re-acceleration in the bulk, assuming Kolmogorov diffusion in the bulk (left) or Kraichnan diffusion (right). The dashed lines show the result in the case of a large shock filling factor $w=0.3$, while the solid lines show the result in the more realistic case $w=0.1$. The various colors compare the effect of decreasing the size of the transition layer.}
   \label{fig:result_reacc}
 \end{figure*}

Our model is sketched on Fig.~\ref{fig:scheme_surfaces}. We consider a wedge of the shock which extends in the azimuthal direction over an angle $\Delta\psi$. The bulk flow is taken to be predominantly in the $\Delta\psi$ direction. The inner surface at radius $r = R_s$ is the WTS surface. The outer surface at $r=R_p$ is the (momentum dependent) escape surface which is defined as follows. We introduce the diffusion length of the particles $L_d$ via:
\begin{equation}
    \int_{R_s}^{R_s+L_d} \dd r' \, \frac{u_r}{D} = \xi
    \, ,
\end{equation}
where $\xi = \mathcal{O}(1)$ is a measure of the confinement. If $L_d < l_s$, then the particles are screened by the radial outflow in the transition layer and the escape boundary is set at $R_p = R_s+l_s$, where the particles are advected in the azimuthal bulk flow. If $l_s<L_d<R_s$, then particles have a non-negligible probability to diffuse back to the shock surface, even though their transport can be influenced by the azimuthal bulk flow. We argue that particles which are within a diffusion length from the shock surface are likely to be further accelerated, while particles beyond a diffusion length are advected in the bulk and leave the accelerator. In these energy bands, the boundary $R_p$ is therefore set at a diffusion length beyond the shock surface: $R_p = R_s+L_d$.
Finally, particles whose momentum is such that $L_d \geq R_s$ leave the accelerator since this is the condition constraining the maximum energy at a spherical shock \citep{morlino2021}. We therefore set $R_p= 2 R_s$ for these particles.
Beyond the escape surface, particles are mixed in the bulk and we therefore have the boundary condition $f(R_p) = \bar{f}$, where $\bar{f}$ is the distribution function averaged over the turbulence in the bulk.

We additionally assume that, beyond the transition layer ($r > R_s+l_s$), an azimuthal flow enters and leaves through the side surfaces with velocity $u_\perp$. This setup maximises the chances that bulk particles reach the accelerator by being advected in the azimuthal direction and then diffusing in the radial direction.

\subsection{Solving the transport equation}
Because the system is locally spherically symmetric up to the radius $R_s+l_s$, the solution known in the case of an isolated WTS in spherical symmetry holds upstream and around the WTS. A detailed study of this solution was done in \citet{morlino2021}. The solution upstream of the WTS reads:
\begin{equation}\label{TE_reacc_upstream}
    D \d_r f - u f = \frac{G_1}{r^2} \, .
\end{equation}
The function $G_1$ defined in \citet{morlino2021} accounts for adiabatic losses in the cold stellar wind and introduces an exponential cut-off at the momentum such that $D_1 \approx u_1 R_s$, where $D_1$ and $u_1$ are respectively the diffusion coefficient and wind velocity upstream of the WTS. If the diffusion is not dramatically suppressed downstream of the WTS, the downstream limitation on the maximum energy, $D_2 \lesssim u_2 R_s$, is more stringent, and the RHS in Eq.~\ref{TE_reacc_upstream} can be ignored.

\bigbreak
Integrating the transport equation around the WTS then gives:
\begin{equation}\label{TE_integratedaroundshock}
    \frac{-\phi_{WTS}}{2 \Delta\psi R_s^2} - u_2 f_s + \left[ D \d_r f \right]_{R_s^-} = - \Delta u \frac{p}{3} \d_p f_s + Q_0 \delta(p-p_i) \, ,
\end{equation}
where $\phi_{WTS}$ is the flux of particles downstream of the WTS, $\Delta u$ is the velocity jump at the shock, $\Delta \psi$ is the angle over which the shock extends in the azimuthal direction, and the injection term is normalised such that $Q_0=  \eta n_1 u_1/4 \pi p_i^2$. While the gradient of the distribution function upstream of the shock is readily obtained from Eq.~\ref{TE_reacc_upstream}, the gradient downstream of the shock must be calculated by solving the transport equation downstream of the shock, accounting for the incoming and escape fluxes through the four surfaces defined in Fig.~\ref{fig:scheme_surfaces}. We make the simplifying assumption that $\nabla \cdot \v{u} \approx 0$ in the hot medium downstream. Then the flux of particles is conserved in this region:
\begin{equation}\label{flux_cons_downstream}
    \oiint \( \v{u} f - D \nabla f \) \cdot \v{\dd S} = 0 \, .
\end{equation}

The integral can be evaluated on four surfaces which enclose the downstream volume: the WTS surface, a spherical escape boundary at $r = R_p$, the side boundary surface $S_1$ from which the azimuthal flow enters, and the side boundary surface $S_e$ from which the azimuthal flow leaves. The transport in the non-radial direction is advection-dominated such that we can assume $\d_\theta f , \, \d_\phi f \ll r \d_r f$. The azimuthal fluxes respectively through the surfaces $S_1$ and $S_e$ are then:
\begin{align} \label{sidefluxes}
    \phi_1 \approx - S_1 \bar{u}_\perp \Bar{f}
    \, , \quad 
    \phi_e \approx S_e \bar{u}_\perp f_s = S_1 \bar{u}_\perp f_s
    \, , 
\end{align}
where $\bar{u}_\perp = \int \dd \theta \int_{R_s}^{R_p} \dd r \, r u_\perp/S_1$. Furthermore, neglecting the gradient in the non-radial directions simplifies the transport equation to:
\begin{equation}
    r^2 u_r \d_r f - D \d_r r^2 \d_r f = 0 \, .
\end{equation}
Using the boundary conditions $f(r=R_s) = f_s$ and $f(r=R_p) = \bar{f}$, we obtain:
\begin{equation}\label{general_implicit_phip}
    f = f_s \exp\( \int_{R_s}^{r} \dd r' \frac{u_r}{D} \) + \frac{\bar{f} - f_s  e^{\alpha_p}}{1-e^{\alpha_p}} \( 1- \exp\( \int_{R_s}^{r} \dd r' \frac{u_r}{D} \) \)
    \, ,
\end{equation}
where $\alpha_p = \int_{R_s}^{R_p} \dd r' \frac{u_r}{D}$.

The flux at the boundary $R_p$ is therefore:
\begin{align} \label{fluxp}
    \phi_p = 2 \Delta\psi R_p^2 u_p \frac{\bar{f} - e^{\alpha_p} f_s}{1-e^{\alpha_p}} \, ,
\end{align}
where $u_p = u_r(R_p)$. We identify the escape flux $\propto f_s$ and the flux of incoming bulk particles $\propto \bar{f}$ which can be re-accelerated. Additionally, bulk particles can also enter or leave the accelerator via the side surfaces, which is encoded in the fluxes $\phi_1$ and $\phi_e$ given by Eq.~\ref{sidefluxes}.

We can express the flux downstream of the WTS as function of $\bar{f}$ and $f_s$:
\begin{multline}
    \phi_{WTS} =  - \phi_e - \phi_p - \phi_1 
    \\
    = \bar{u}_\perp \( R_{p}^2 - (R_s+l_s)^2 \)   \( \bar{f} - f_s \)
    + 2 \Delta\psi R_p^2  u_p \frac{\bar{f} - f_s e^{\alpha_p}}{e^{\alpha_p} - 1} \, .
\end{multline}
The above expression can be used in Eq.~\ref{TE_integratedaroundshock} to solve for $f_s$ around the shock. This can be done for any prescription of the velocity profile, although for simplicity we assume $u_r \propto 1/r^2$. This leads to:
\begin{multline}\label{TEatshockworkedoutFINALLYENDOFNIGHTMARE}
    p \d_p f_s + s f_s
    + f_s \(
    \frac{3 u_2/\Delta u}{e^{\alpha_p}-1} 
    + \frac{\beta}{2 \xi} \( \(\frac{R_p}{R_s} \)^2 - 1 \)
    \)
    \\
     = \frac{\beta}{2 \xi} \frac{R_{p}^2 - (R_s+l_s)^2}{R_s^2}  \bar{f} 
     + \frac{s}{\sigma}  \frac{\bar{f}}{e^{\alpha_p} - 1} 
     +  \frac{3}{\Delta u} Q_0 \delta(p-p_i) \, ,
\end{multline}
where $s = 3 u_1/\Delta u$ and $\beta = 3 \xi \bar{u}_\perp/(\Delta u \Delta \psi )$. The two additional terms appearing on the LHS produce two sources for spectral cut-offs, corresponding to the escape through the surfaces $\{r=R_p\}$ and $S_e$. The integration of Eq.~\ref{TEatshockworkedoutFINALLYENDOFNIGHTMARE} is straightforward and we eventually obtain the spectrum of accelerated and re-accelerated particles at the shock, assuming for simplicity that Bohm diffusion applies:


%
\begin{multline}\label{solution_reacceleration_shock}
    f_s(p) = \int^p \frac{\dd p_0}{p_0} \(\frac{p}{p_0}\)^{-s}
    e^{- \frac{6D}{(\sigma-1)u_2 R_s}}
    e^{- \frac{\beta D}{u_2 R_s}  }
    \\ \left\{
    \frac{R_{p_0}^2 - (R_s+l_s)^2}{2 \xi R_s^2}  \beta \bar{f} (p_0)
     + \frac{s}{\sigma}  \frac{\bar{f}(p_0)}{e^{\alpha_{p_0}} - 1} 
     +  \frac{3 Q_0}{\Delta u} \delta(p_0-p_i) \right\} \, .
\end{multline}
The total flux at the boundary of the accelerator is obtained by summing Eqs.~\ref{sidefluxes} and~\ref{fluxp}: $\phi_s = \phi_1 + \phi_e + \phi_p$. This flux enters as a source term in Eq.~\ref{TEbulkaveraged}. The bulk distribution function is now averaged over the core, such that the transport equation in the bulk reads:
\begin{multline}\label{TEbulk}
    \d_t f_c + \frac{f_c}{\tau_e} =  \frac{4 \bar{u}_c}{3 R_c} p \d_p f_c + \frac{1}{p^2} \d_p p^2 D_p \d_p f_c 
    \\
    + \frac{3 w u_w }{\sigma R_s}
    \( 
   \frac{R_{p}^2 - (R_s+l_s)^2}{R_s^2} \frac{ (\sigma-1) \beta}{6  \xi}
    \( f_s - \bar{f} \)
    + \frac{ e^{\alpha_p} f_s - \bar{f}}{e^{\alpha_p}-1}
    \)
    \, ,
\end{multline}
where $w = N_* (R_s/R_c)^3 \Delta \psi/(2\pi)$ is the volume filling factor of the wind cavities.

Eq.~\ref{solution_reacceleration_shock} does not depend on space therefore the volume average is trivial and we can calculate $f_s$ as function of $f_c$ by setting $\bar{f} = f_c$. Therefore Eq.~\ref{TEbulk} becomes an integro-differential equation for $f_c$ which can be solved numerically, similarly to the expressions obtained in \citet{bykov1990,klepach2000}. The two key parameters controlling the re-acceleration rates are $\beta$ and $w$. Typically we expect $\bar{u}_\perp \sim \bar{u}_c \approx 0.14 u_w$ (the azimuthal velocity is of the order of the bulk flow velocity as seen in the simulation) and therefore $\beta \sim 1/\Delta\psi$. In the following we fix $\beta=1$, i.e. the azimuthal flow extends over a length comparable to the shock radius ($\Delta \psi = 1$).

Fig.~\ref{fig:result_reacc} shows the resulting spectra for various values of the transition layer $l_s$. A hard component is only obtained if $l_s$ is smaller than a few percent of the size of the shock (yellow and blue lines), which, according to our simulations, is not expected to occur in a realistic scenario. Furthermore, the hardening is only pronounced if the volume filling factor of the shocks is large ($w=0.3$, dotted lines), which is only expected in the most massive compact stellar clusters. Overall, we conclude that shock re-acceleration produces hard signatures at high energies only in a restricted parameter space which is likely never achieved in reality.

\section{On the absence of collective acceleration effects in a system of colliding winds}\label{sec:CollidingWinds}
In this section we revisit the possibility that particles accelerated in a system of stationary colliding flows develop a hard distribution function ($f \propto p^{-3}$ at high energies according to \citealt{bykov2013}). This result was obtained in a one-dimensional stationary model which needs refinement, in particular because in such a setup the outflow escaping from the collision region is not modelled hence the mass, energy and momentum are not conserved.

\subsection{Modelling colliding winds}

\begin{figure}
          \centering
              \includegraphics[width=1\linewidth]{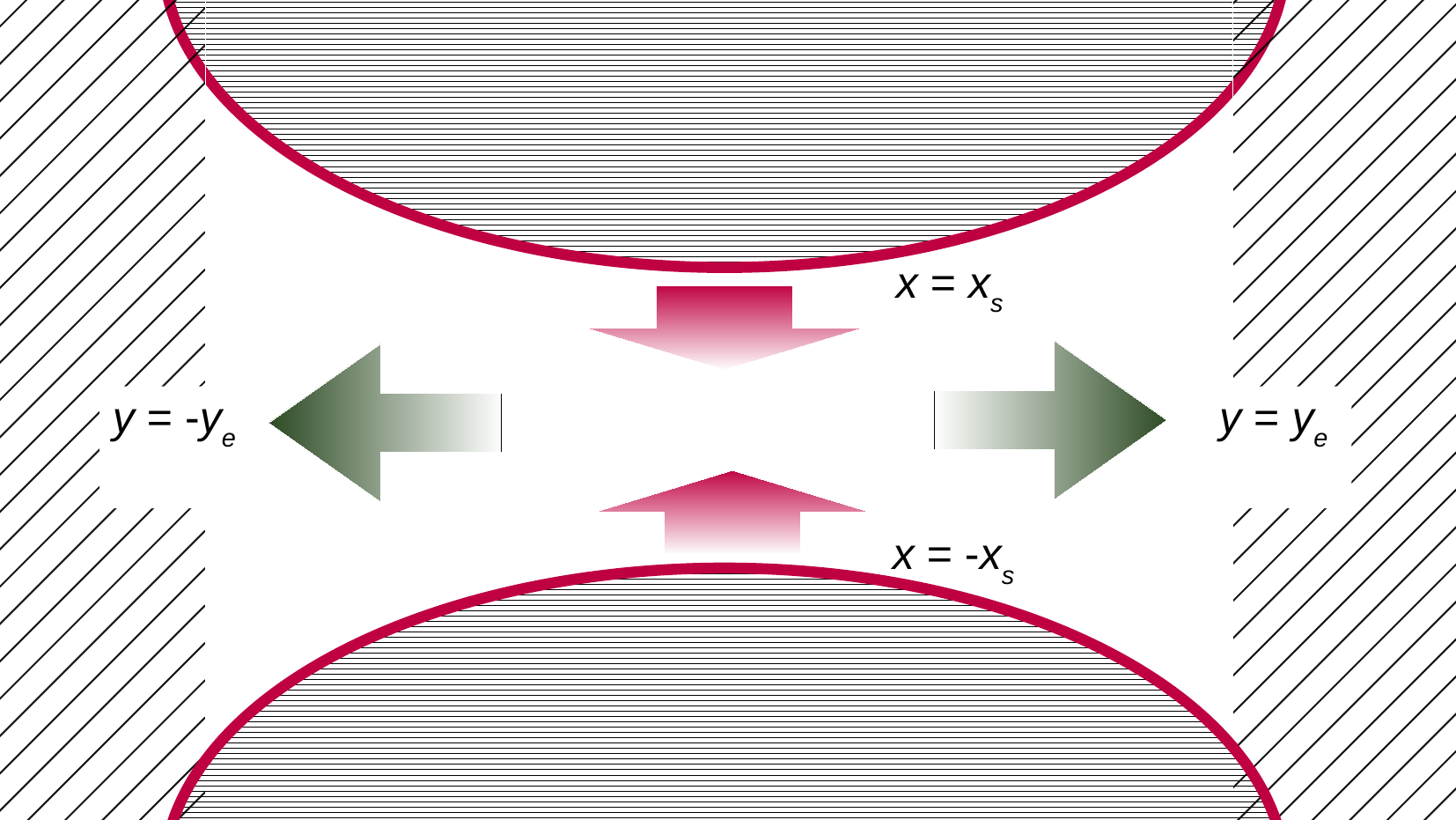}
   \caption{Scheme of the colliding winds. The stationary shocks are located at $x= \pm x_s$. The orthogonal outflow escapes in the bulk at $y,z = \pm y_e$. We solve the transport equation in Cartesian geometry assuming a linear velocity profile in all directions.}
   \label{fig:scheme_collision}
 \end{figure}

The bottom left panel of Fig.~\ref{fig:simulation_zoomed} shows the flow configuration in a stationary system of colliding winds, which is also sketched in Fig.~\ref{fig:scheme_collision}. The plasma is decelerated linearly in the direction parallel to the collision axis ($x$-axis in the following), and accelerated linearly in the orthogonal plane: 
$u_x = - u_s x /x_s$, $u_y = u_e y /y_e$, $u_z = u_e z /y_e$, where $u_s$ is the velocity downstream of either shock, $x_s$ is the distance of either shock to the plane of symmetry (see Fig.~\ref{fig:scheme_collision}). The colliding flows merge smoothly across the plane of symmetry, along which $u_x = 0$ (there is no contact discontinuity). Along the orthogonal direction, the linear increase of the outflow stops at a distance $y_e \approx 0.1 - 0.2$~pc from the centre of the system, which is typically the extension of the shocks in the orthogonal direction. Beyond the distance $y_e$, the flow is contaminated by the bulk flows and in the following we will therefore set a free escape boundary for the particles at $y = \pm y_e$. Assuming that the geometry is approximately Cartesian and that the downstream flow is incompressible (as seen on the middle panel of Fig.~\ref{fig:velocitydivmap}, $\nabla \cdot \v{u} \approx 0$ close to the shocks), we have $u_s/x_s \approx 2 u_e/y_e$.

We emphasise that accounting for a realistic flow configuration in the downstream region with a stagnation plane and an orthogonal outflow is crucial, a fact which was neglected in previous works \citep{bykov2013,reimer2006,malkov2023}.

\subsection{Solution of the transport equation}

The transport equation upstream of either shock along the axis orthogonal to the shock planes is taken to be identical to the case of an isolated planar shock, with a constant flow in the direction orthogonal to the shock. Denoting the upstream velocity $u_1$, we have $D_x \d_x f = - u_1 f$ upstream, such that the transport equation integrated around the shock leads to:
\begin{equation}\label{TEcollidingaroundshock}
    p \d_p f_s = - s f_s - \frac{3 D_x}{\Delta u} \left. \d_x f \right|_{d} \, ,
\end{equation}
where $s = 3 u_1/\Delta u$, $\Delta u = u_1-u_s$, $u_s$ is the velocity immediately downstream of the shock along the $x$-axis, $\left. \d_x f \right|_{d}$ is the derivative of the distribution function immediately downstream of the shock. The latter is worked out by solving the transport equation in the common downstream region.

To solve the transport equation in the downstream region we use the method of separation of variables. We write the distribution of particles $f(x,y,z) = g(x)h(y)k(z)$ denoting $x$ the direction orthogonal to the shock planes. By symmetry, $h=k$. The equations for $g$ and $h$ then read:
\begin{align}
    - x_s^2 D \d_x^2 g - x \d_x g = - \frac{C}{2} g \, , \label{TEforg}
    \\
    - y_e^2 D_\perp \d_y^2 h + y \d_y h = C h \, , \label{TEforh}
\end{align}
where $D = D_x/(u_s x_s)$ is the dimensionless diffusion coefficient along the $x$ axis and $D_\perp = D_y/(u_e y_e) = D_z/(u_e y_e)$ is the dimensionless diffusion coefficient along the orthogonal direction. $C$ is a constant in the domain. Multiplying Eq.~\ref{TEforg} by $u_s/x_s$, the RHS becomes $-(u_s/x_s) (C/2) g = -(u_e/y_e) C g$, which shows that this term can be interpreted as an escape term in the orthogonal direction due to the advection of the particles with a velocity of order $u_e$. Physically, we therefore expect $ 0 \leq C \leq 1$. Note that for low energy particles, $D_\perp \ll 1$, the only value of $C$ which provides a physical solution with a reflective boundary at the origin, $\left. \d_y h \right|_{y=0} = 0$, is $C=0$, which implies $\d_x g = 0$ and $f\propto p^{-s}$, as expected: low energy particles only probe a small region downstream of either shocks and are not affected by the presence of the other shock.

The solution to Eq.~\ref{TEforh} with a reflective boundary at $y=0$ is the confluent hypergeometric function:
\begin{equation}
    h(y) \propto ~_1F_1 \( -\frac{C}{2} , \frac{1}{2} , \frac{1}{2 D_\perp} \frac{y^2}{y_e^2} \)  \, .
\end{equation}
Imposing a free escape boundary at $y=y_e$ then leads to an implicit equation for the constant $C$:
\begin{equation}
    ~_1F_1 \( -\frac{C}{2} , \frac{1}{2} , \frac{1}{2 D_\perp} \) = 0 \, ,
\end{equation}
which can be solved numerically. We find $C\ll 1$ for $D_\perp \ll 1$, as expected, and $C \propto D_\perp$ otherwise.

\begin{figure}
          \centering
              \includegraphics[width=1\linewidth]{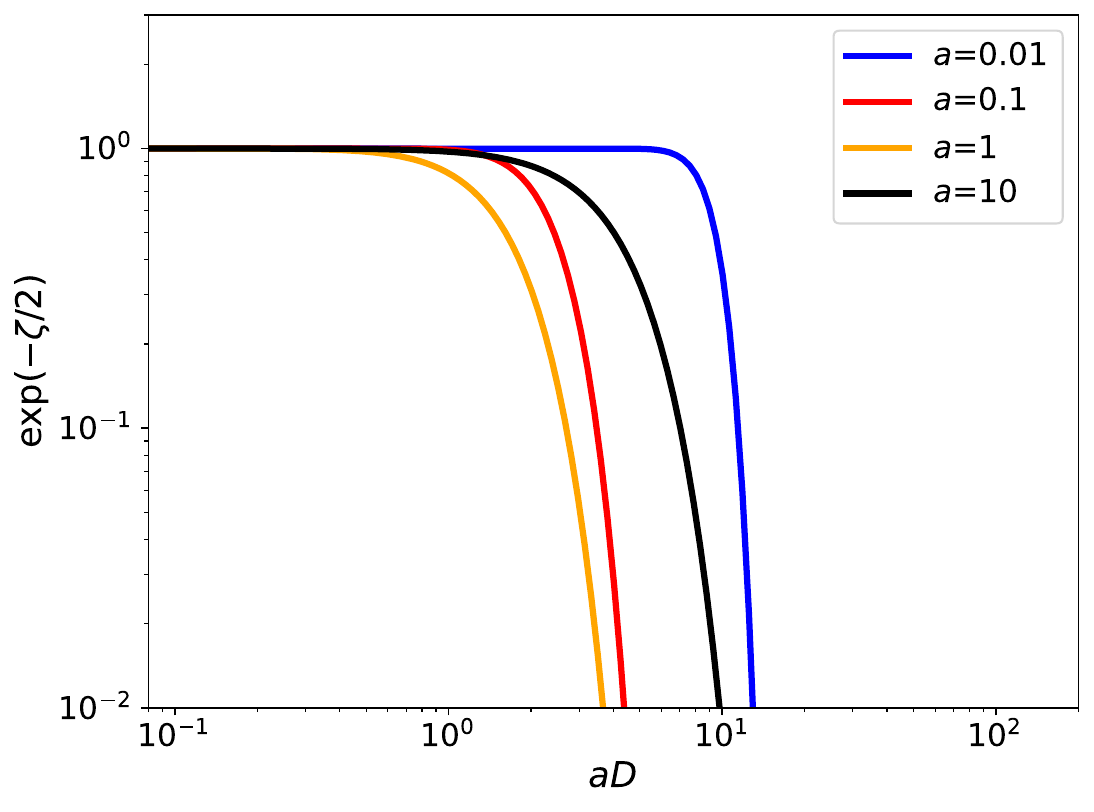}
   \caption{Distribution of particles accelerated in a system of stationary colliding shocks (solution to Eq.~\ref{equationforzeta}, the power-law being factored out), as function of the diffusion coefficient $D = D_x/(u_s x_s)$ and $a = \sqrt{D_\perp/D}$.}
   \label{fig:colliding_shocks_cutoff}
 \end{figure}

The solution to Eq.~\ref{TEforg} for the function $g$ is also a hypergeometric function, and thus:
\begin{equation}
    \left. \d_x g \right|_{x=x_s^-} = g(x_s) \frac{C}{2 D x_s} 
    \frac{~_1F_1 \( 1 - \frac{C}{4} , \frac{3}{2} , -\frac{1}{2 D} \)}{~_1F_1 \( - \frac{C}{4} , \frac{1}{2} , -\frac{1}{2 D} \)}
    \, .
\end{equation}
Let us now define a function $\zeta$ such that $f_s(p) \equiv p^{-s} \exp{\(-\frac{3 u_2}{2\Delta u}\zeta(p)\)}$. Assuming that the diffusion coefficient is a power-law with respect to momentum: $D \propto p^q$, Eq~\ref{TEcollidingaroundshock} provides:
\begin{equation}\label{equationforzeta}
    D \d_D \zeta = \frac{C}{q}
    \frac{~_1F_1 \( 1 - \frac{C}{4} , \frac{3}{2} , -\frac{1}{2 D} \)}{~_1F_1 \( - \frac{C}{4} , \frac{1}{2} , -\frac{1}{2 D} \)} 
    \, .
\end{equation}
One can show that the RHS is always positive, therefore $\zeta$ is always positive as well and the spectrum at the shock $f_s(p)$ will never be harder than the single shock solution $p^{-s}$. Fig.~\ref{fig:colliding_shocks_cutoff} shows the result of the numerical integration of the function $\zeta$ for various values of $a\equiv \sqrt{D_\perp/D}$ (which, remarkably, is the only parameter of the problem). 

The result is a sharp cut-off around $D \propto 1/a$, i.e. the maximum momentum is such that $D_x D_y \sim (u_s y_e)^2$. This is reminiscent of the Hillas criterion. Low energy particles are efficiently accelerated in the vicinity of either shock until they propagate beyond a diffusion length downstream, in which case they escape via advection, first along the parallel axis, then along the orthogonal direction. On the other hand, particles above the maximum energy threshold are lost at the free escape boundary. Even particles whose parallel diffusion length is much larger than the distance between the shocks, $D_x \gg u_s x_s$, do not experience any collective acceleration mechanism. Although this might at first seem counter-intuitive, it is in fact expected. It follows since the flow decelerates along the parallel axis and the parallel velocity is effectively zero along the plane of symmetry, which implies that the distribution of particles is always isotropised in this region, in all energy bands. Then, at momenta such that the parallel diffusion length is larger than $x_s$, the particles can diffuse back to either shock, exactly as they would in the case of a single shock. Note that this is a general argument which holds whatever geometry and flow configuration is assumed. A system of colliding winds is in fact equivalent to a single shock with an outflow bent downstream. The latter does not impact the diffusive shock acceleration mechanism.

\subsection{The impossible regime $\Lambda > 2 x_s$}

The only way to trigger a collective process is to prevent isotropisation in the downstream region. This is only the case if the mean free path of the particles is larger than the distance between the shocks: $\Lambda \sim D_x/v > 2 x_s$. In this regime, the diffusion approximation breaks down in the region between the colliding shocks, such that the usual transport equation should not be used. One can still consider, as an upper bound estimate, that the propagation is ballistic in between the shocks, in which case the downstream region can be assumed to be infinitesimally small and the transport equation can be integrated around both shocks simultaneously: $(2 u_1/3) p\d_p f_s = 2 D_0 \d_x f_s = - 2 u_1 f_s$, which immediately provides the hard solution $f \propto p^{-3}$ of \citet{bykov2013}.

However we argue that the regime $\Lambda > 2 x_s$ is never reached in a realistic system. Indeed, this criterion implies $D_x > 2 x_s v$. On the other hand, according to the maximum energy criterion derived above, we have $D_x D_y < (u_s y_e)^2$. As shown earlier, $2x_s \approx y_e u_s/u_e$, implying $u_e > \sqrt{D_y/D_x} v$. Since $v \approx c$, this inequality can only be satisfied if the diffusion is strongly suppressed in the orthogonal direction. It is unclear which physical mechanism could cause such a suppression, especially since the magnetic field lines are expected to be either randomised by downstream instabilities, or compressed along the shock surfaces and sheared along direction of outflow. This should in fact lead to $D_x \ll D_y$. The physical interpretation of the above calculation is straightforward: particles escape the system when reaching the Hillas limit $D_y > u_e y_e$, which happens at an energy much lower than that would be required to get $\Lambda > 2 x_s$.

\bigbreak
It is therefore impossible to develop a hard spectral tail in a system of colliding shocks when a realistic flow configuration is considered. The same conclusion holds in a time-dependent collision, as shown in \citet{vieu2020}. Although in this case the shocks can come  close together, there is too little time for collective effects to develop. Finally, our argument also applies to colliding winds in binary systems, which in any case would be too small to confine accelerated particles with energies beyond several TeV \citep{eichler1993}, even in the extreme case of $\eta$~Carinae \citep{etaCarinae2020}.

\section{Damping of stochastic re-acceleration in the bulk}\label{sec:StochasticReacc}
Collective re-acceleration effects may also arise when the particles propagate in a medium characterised by strong turbulence. According to quasi-linear theory, the spatial diffusion is ineluctably accompanied by a diffusion in momentum. The diffusion coefficients can be calculated from quasi-linear theory as \citep[e.g.][Eq.~2.57]{vieuphd}:
\begin{align}
    D &= \frac{v}{3} L_T \( \frac{r_L}{L_T} \)^q \, ,
    \\
    D_p &= \frac{v_A^2 p^2}{\alpha_\nu D} \, , \quad \alpha_\nu = (8+\nu^2-6\nu)(\nu+2)\nu \, , \label{EqDpfactor2neglected}
\end{align}
where $v$ is the particle velocity, $L_T$ is the injection scale of the turbulence, $r_L$ is the Larmor radius of the particles, $q=2-\nu$ with $\nu$ the spectral index of the turbulence, and $v_A$ is the Alfv\'en velocity in the plasma.
As discussed in Section.~\ref{sec:transportinbulk}, the correction due to strong compressive modes brings at most a factor 2 in Eq.~\ref{EqDpfactor2neglected} when $v_A \approx \delta u$, with $\delta u$ the characteristic velocity of the turbulence. For simplicity we neglect this small correction in the following. From the diffusion coefficients we can derive the re-acceleration timescale: $\tau_p = p^2/(2 D_p)$, as well as the escape time: $\tau_e \approx \(\frac{4 \bar{u}_c}{R_c} + \frac{8 D}{ R_c^2} \)^{-1}$ (Eq.~\ref{escapetime}), 
with $\bar{u}_c$ the mean radial velocity in the core and $R_c$ the radius of the core. 

\begin{figure}
    \centering
    \includegraphics[width=\linewidth]{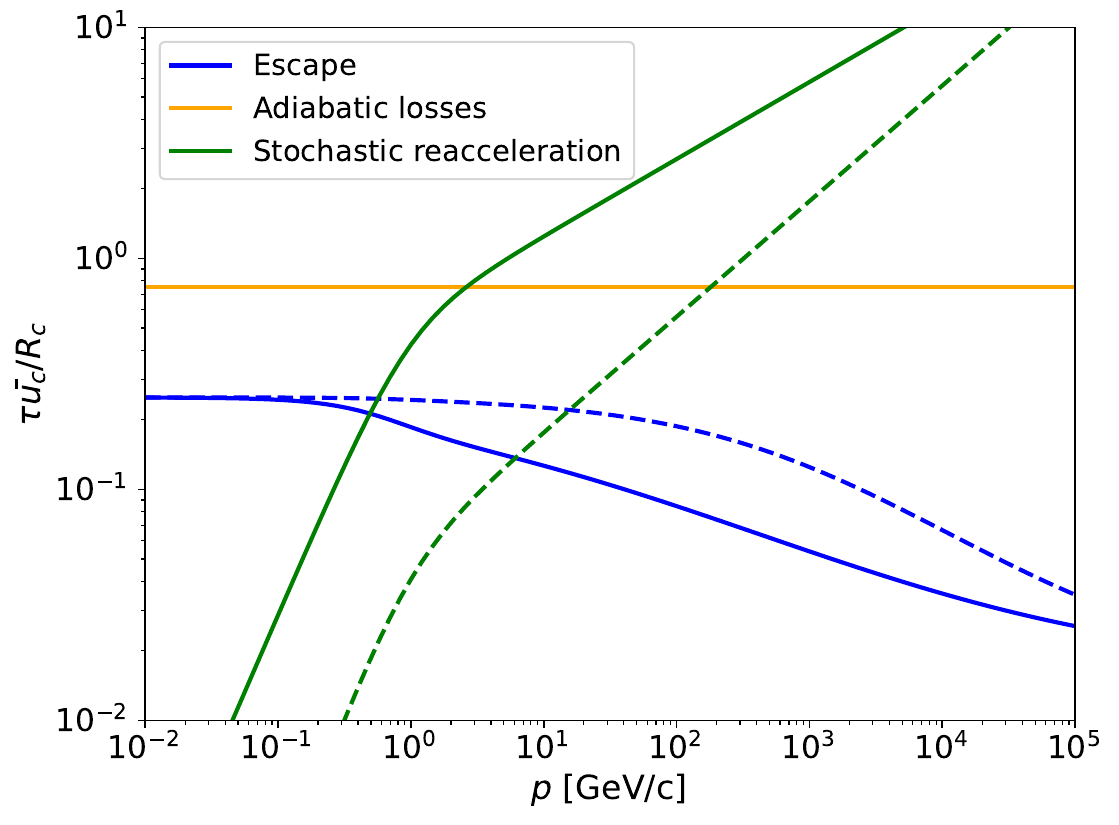}
    \caption{Stochastic re-acceleration timescale in the core, compared to the escape time, for $\bar{u}_c = \delta u = v_A = 200$~km/s, $L_T = 0.1$~pc, $R_c = 1$~pc, $B = 86$~\textmu G. The solid line shows the case of Kolmogorov turbulence ($\nu = 5/3$) while the dashed line shows the case of Kraichnan turbulence ($\nu = 3/2$).}
    \label{fig:timescalescoreDp}
\end{figure}

Fig.~\ref{fig:timescalescoreDp} shows a comparison between these timescales. Although stochastic re-acceleration is a second order process, one can see that the turbulence is in principle strong enough to allow efficient re-acceleration up to the GeV band and one could naively expect the spectra to harden noticeably. One also notices that adiabatic losses are only expected to have a mild impact on the results. Let us therefore start by considering the averaged transport equation in the bulk, without losses and with diffusion in momentum:
\begin{equation}\label{TE_stochastic_reacc}
     \frac{f_c}{\tau_e} = \frac{1}{p^2} \d_p p^2 D_p \d_p f_c + Q(p)
     \, ,
\end{equation}
where the source term $Q(p) \propto p^{-s} e^{-p/p_m}$ models the injection at the interfaces with the wind cavities.

\subsection{Stochastic Acceleration - Analytic solution}
\begin{figure}
    \centering
    \includegraphics[width=\linewidth]{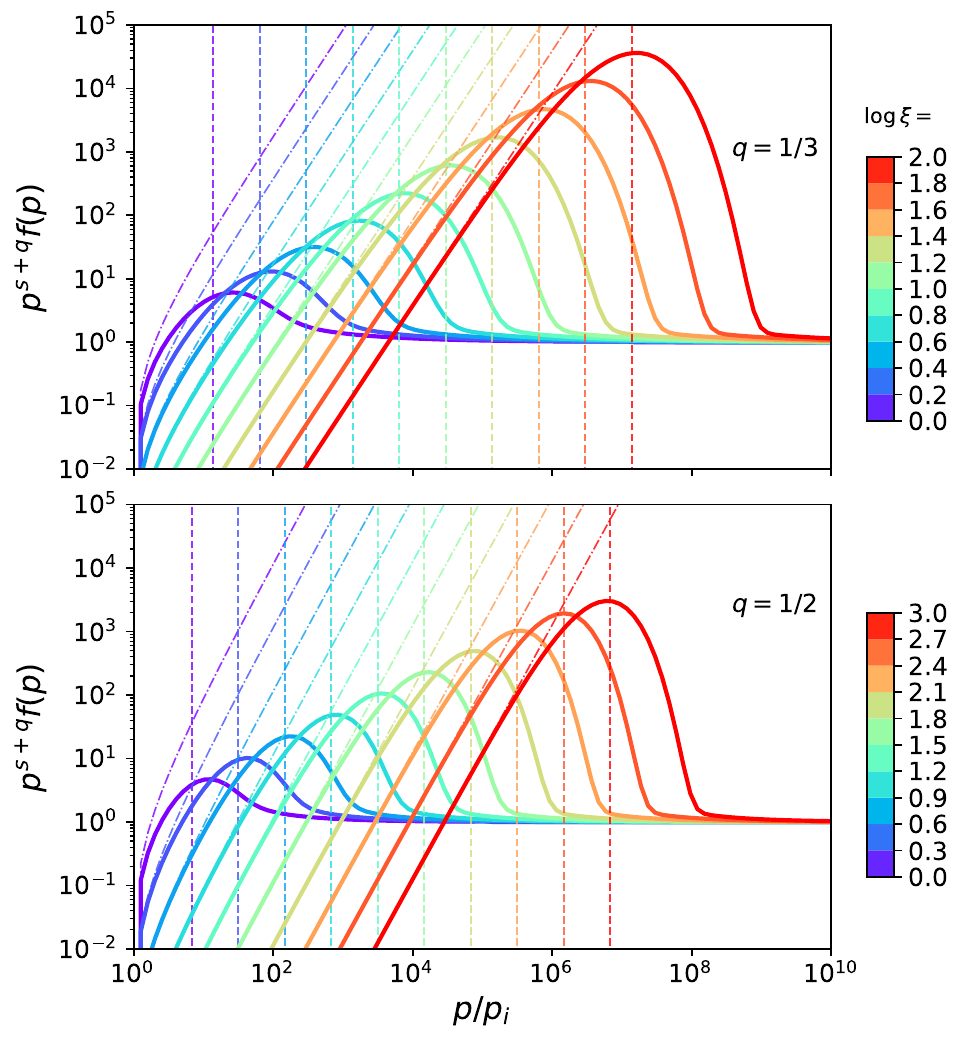}
    \caption{Each plot shows the solution of Eq.~\ref{TE_stochastic_reacc} for various values of $\xi = \sqrt{\tau_0/(2\tau_p)}$ (shown by the colorbar). The solutions computed by numerical integration of Eq.~\ref{finalresultf} are displayed by the solid lines. The dashed-dotted lines show the approximation of the solution at low momenta (Eq.~\ref{fDplowpapprox}). The vertical dotted lines show the position of the bump as given by Eq.~\ref{bumpposition}. The top panel shows the case of Kolmogorov turbulence while the bottom panel shows the case of Kraichnan turbulence.}
    \label{fig:compute_all_eta}
\end{figure}

In the case where the transport coefficients have power-laws dependence on the momentum, $\tau_e = \tau_0 (p/p_i)^{-\delta}$, $D_p = p^2/(2\tau_p) (p/p_i)^{-q}$, Eq.~\ref{TE_stochastic_reacc} can be solved analytically. The derivation is shown in Appendix~\ref{app:AnalyticStochastic}. The semi-analytical solution given by Eq.~\ref{finalresultf} is displayed in Fig.~\ref{fig:compute_all_eta} in the case of a power-law injection, for various values of the parameter $\xi = \sqrt{\tau_0/(2\tau_p)}$.

The re-acceleration mechanism hardens the low-energy part of the spectrum. Since the process is less and less efficient, the particle distribution function piles up at some breaking point. The spectrum then turns over, and must steepen in order to recover the high-energy solution, which is simply the injection spectrum modulated by the escape ($f(p) \propto p^{-q-s}$). A bump is therefore expected to develop in the GeV band: this is a typical signature of stochastic re-acceleration. When the re-acceleration efficiency $\xi$ increases, the amplitude of the bump increases, and its peak is shifted to higher momenta.

One can show that between the injection and the peak of the bump, the spectrum is well approximated by a power-law, $f(p) \propto p^{q-3}$ (see Eq.~\ref{fDplowpapprox}). In the case of hydrodynamic turbulence, $q=0$, we retrieve the well-known result $f(p) \propto p^{-3}$. Flatter turbulence spectra produce even harder low-energy components.

Assuming $\delta u = v_A$, we see in Fig.~\ref{fig:timescalescoreDp} that in the regime of efficient re-acceleration, $\tau_p < \tau_e$, the spatial transport is dominated by advection. The escape time relevant to calculate the parameter $\xi$ is then the advection time: $\tau_e \approx R_c/(4 \bar{u}_c)$. The parameter $\xi$ is then obtained as:
\begin{equation}
    \xi \approx \sqrt{ \frac{3}{8  \alpha_\nu } \frac{v_A^2}{ \bar{u}_c c } \frac{R_c}{L_T} \( \frac{L_T}{r_{L,i}} \)^q  } \, ,
\end{equation}
where $r_{L,i}$ is the Larmor radius at the injection momentum. Typically, for $v_A \approx \bar{u}_c \sim 0.001 c$, $R_c = 10 L_T = 1$~pc, $B = 100$~\textmu G, $p_i = 0.01$~GeV, we expect $\xi \sim \mathcal{O}(1)$.

One concludes that a Kolmogorov turbulence spectrum is only expected to produce a noticeable re-acceleration bump with a peak around 0.1 - 10~GeV, similarly to the result obtained in \citet{vieu2022} for the transport in the superbubble beyond the cluster core. In the case of Kraichnan turbulence, the bump can in principle be shifted up to TeV energies, although care must be taken at this stage as a considerable amount of energy is transferred from the turbulence to the particles. The test-particle approach is not expected to provide reliable results in this case, as the turbulence cannot be considered as an infinite reservoir of energy anymore and its damping by the particle feedback must be taken into account. This will dramatically decrease the re-acceleration efficiency. Including this effect is the purpose of the next section.

\subsection{Non-thermal wave damping}
Assuming that a fraction $\eta$ of the stellar power $P_*$ is transferred to CR acceleration at the shocks, we can estimate the CR energy density in the core as $e_{CR} \approx \eta P_* \frac{R_c}{4 \bar{u}_c}/(4/3 \pi R_c^3)$. On the other hand, the energy density of the turbulence is $e_T = \rho_c \delta u^2 = \rho_c \eta_T u_w^2/16$, where $\eta_T$ is the fraction of the wind kinetic energy transferred into MHD turbulence (see Eq.~\ref{deltauequipartition}). Therefore the CR energy density is in general expected to be much smaller than the energy density of the turbulence:
\begin{equation}
    \frac{e_{CR}}{e_T} \approx 8 \pi^2  \frac{\eta}{\eta_T} \, .
\end{equation}
In this case, the back-reaction of the particles onto the turbulence must be considered. A mean power $(1/p^2)\d_p (v p^2 D_p)$ is continuously transferred via stochastic re-acceleration from the waves to the particles. In virtue of flux conservation, the same power must be subtracted from the turbulence via a damping term. Stochastic re-acceleration becomes self-regulated, with a correction term that reads, in the case of a Kolmogorov cascade \citep{vieu2022}:
\begin{equation}\label{DpfeedbackKol}
    D_p \approx \( \frac{L_T}{r_L} \)^{\frac{1}{3}} \frac{p^2 v_A^2}{v L_T} \( 1 - \frac{C_{Ko}}{v_A \rho} \( \frac{L_T p}{r_L} \)^{\frac{1}{3}} \int_p \dd p' p'^{\frac{8}{3}} f(p')  \)^2
    \, ,
\end{equation}
where $C_{Ko} \approx 3.28$. In the case of a Kraichnan turbulent cascade, the feedback becomes:
\begin{equation}\label{DpfeedbackKra}
    D_p \approx 0.66 \( \frac{L_T}{r_L} \)^{\frac{1}{2}}  \frac{p^2 v_A^2}{v L_T} \( 1 - \frac{C_{Kr}}{v_A \rho} \( \frac{L_T p}{r_L} \)^{\frac{1}{2}} \int_p \dd p' p'^{\frac{5}{2}} f(p')  \)
    \, ,
\end{equation}
where $C_{Kr} \approx 4.65$. The spatial diffusion coefficient is then obtained using the relation $D_p D = v_A^2 p^2/\alpha_\nu$. Eqs.~\ref{DpfeedbackKol},~\ref{DpfeedbackKra} are valid only if the expression in the parentheses is positive. If there exists a momentum $p_*$ such that for $p<p_*$ the term in parentheses becomes negative, it means that the damping is so strong that the turbulence is terminated at wavenumbers $k>1/r_{L,*}$, where $r_{L,*}$ is the Larmor radius at momentum $p_*$. In this case, $D_p = 0$ for any $p\leq p_*$ and an additional prescription must be chosen for the spatial diffusion coefficient. In the following we use the prescription suggested in \citet{vieu2022}:
\begin{equation}
    D \approx \( \frac{\alpha_\nu D_p}{v_A^2 p^2} + \frac{6 \pi}{v L_T} \)^{-1}
    \, ,
\end{equation}
which accounts for the randomness of the background fieldlines when the turbulence is suppressed.

\subsection{Numerical solution}
\begin{figure}
    \centering
    \includegraphics[width=\linewidth]{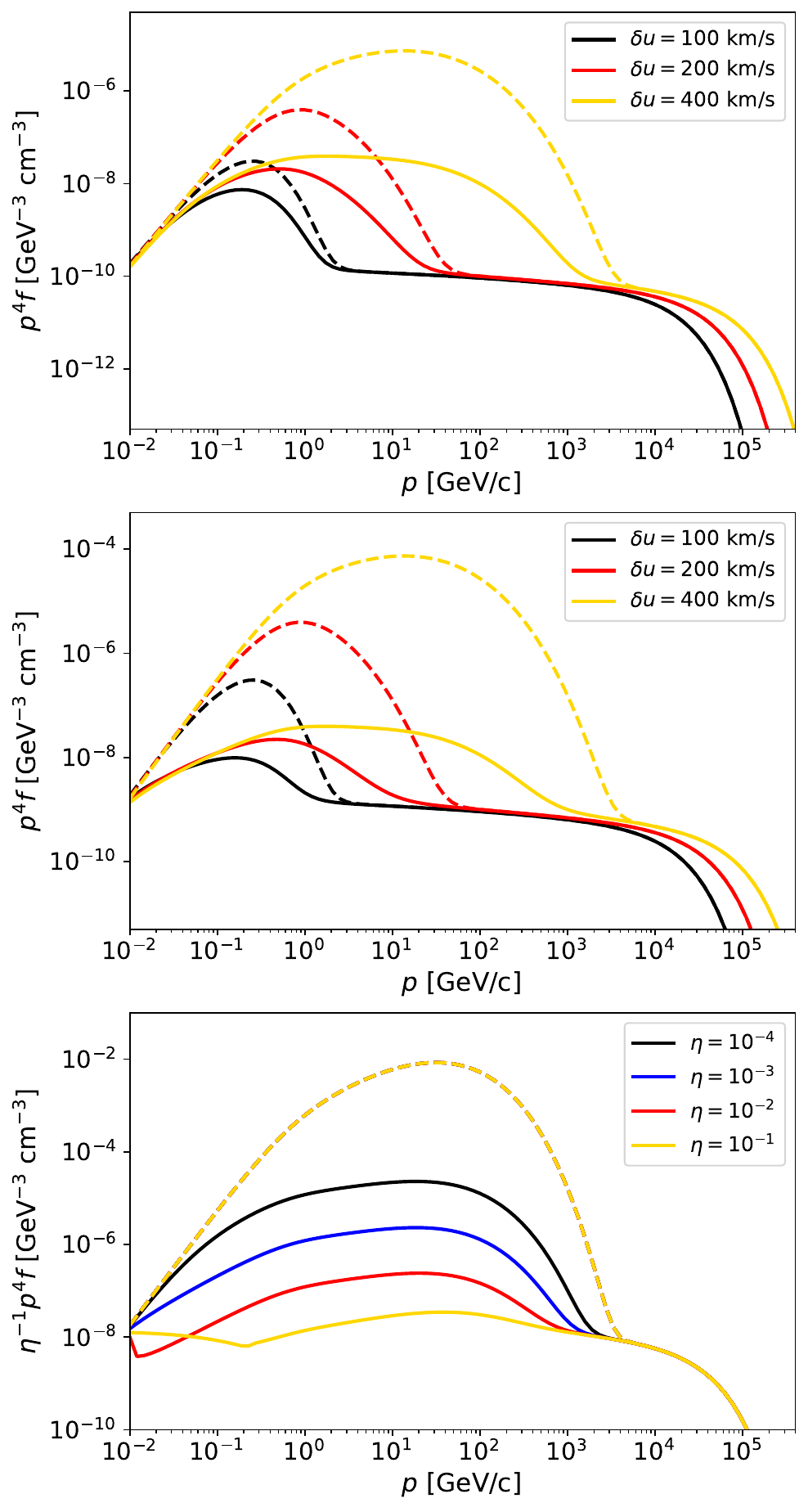}
    \caption{Numerical solution of the transport equation in the bulk, including adiabatic losses and stochastic re-acceleration in the test-particle approximation (dashed lines), compared with the result obtained when accounting for the non-thermal damping of the turbulence (solid lines). Top panel: Kolmogorov turbulence with a CR injection efficiency $\eta = 0.01$. Middle panel: Kolmogorov turbulence with a CR injection efficiency $\eta = 0.1$. Bottom panel: Kraichnan turbulence with $\delta u = 200$~km/s.}
    \label{fig:stochastic_reacc_with_feedback}
\end{figure}

We now solve numerically the transport equation, including adiabatic losses:
\begin{equation}\label{TEbulkaveragedDp}
    \d_t f_c + \frac{f_c}{\tau_e} =  \frac{4 \bar{u}_c}{3 R_c} p \d_p f_c + \frac{1}{p^2} \d_p p^2 D_p \d_p f_c + Q(p) \, ,
\end{equation}
with a power-law injection $Q(p) \propto p^{-4}$. In the following paragraphs we consider the case of protons. We evolve the distribution function using a super-time-stepping algorithm \citep{StSalgo2014} to reach a stationary solution. The diffusion coefficients are renormalised at each super-timestep according to Eqs.~\ref{DpfeedbackKol},~\ref{DpfeedbackKra}. The result is shown in Fig.~\ref{fig:stochastic_reacc_with_feedback} for various values of the strength of the turbulence and the acceleration efficiency.

Already in the case of a Kolmogorov cascade and a small acceleration efficiency ($\eta=1$~\%, top panel), the feedback noticeably reduces the re-acceleration process by decreasing the amplitude of the bump. For an acceleration efficiency of 10~\% (middle panel), the effect of the feedback is dramatic. Even in the case of  strong turbulence, $\delta u = 400$~km/s, the hardening at low energies is  mild. Finally, the case of a Kraichnan turbulence cascade is shown in the bottom panel of Fig.~\ref{fig:stochastic_reacc_with_feedback} with a turbulent velocity $\delta u = 200$~km/s. Even assuming low CR injection efficiencies, one sees that the solution including feedback displays a  mild hardening compared to the test-particle solution. The higher the CR injection efficiency, the stronger the damping. In particular, for an efficiency $\eta = 1$\%, the cascade is terminated at the injection momentum, which produces a small dip as the absence of waves facilitates the escape of low energy particles. In the case of a standard efficiency $\eta = 10$~\%, the cascade is terminated over more than two orders of magnitude and there is almost no re-acceleration. In the realistic range of parameters, stochastic re-acceleration in the cores of stellar clusters is therefore not expected to produce detectable observational signatures.




\section{Conclusions}\label{sec:conclusions}
The cores of massive star clusters are  intricate regions, where potentially hundreds of stellar winds interact in non-trivial ways. This makes them \textit{a priori} promising environments for particle acceleration, although modelling non-thermal phenomena in such systems is a challenging task.

Our goal in this work was to clarify the processes of particle re-acceleration in these environments. A detailed 3D hydrodynamic simulation, resolving individual massive stars in the core, allowed us to investigate the flow configuration of the plasma in the core, which we used as a basis for modelling three mechanisms of particle re-acceleration which have  previously been claimed to produce hard spectra.

We showed that, in contrast to previous claims, multiple shock re-acceleration, shock collisions, and stochastic re-acceleration, do not produce spectral hardening in realistic scenarios, nor do they increase the maximum energy of the particles.

Multiple-shock re-acceleration is inefficient because particles cannot encounter multiple shocks unless their diffusion length is larger than the $\sim 0.01$~pc transition layer which extends downstream of the wind cavities. A hardening can only appear in the highest energy band, and because the wind filling factor is typically small, it is not expected to produce any noticeable feature in gamma-rays.

Particles trapped in a system of colliding winds do not experience a collective acceleration process, because the distribution function is always isotropised in the stagnation plane which separates the shocks, and particles escape in the orthogonal outflow before a regime where the mean free path is larger than the distance between the shocks can be established.

Stochastic re-acceleration is strongly suppressed as the turbulence is damped by the particles themselves, for the non-thermal energy density is generically much larger than the energy density of the turbulence. The diffusion in momentum should not be calculated in the test-particle approximation, otherwise the result violates the conservation of energy. Taking into account the back-reaction of the particles onto the MHD waves leads to a dramatic suppression of the hard component at low energies, and the remaining,  mild, hardening is not expected to produce detectable observational signatures.

In conclusion, none of these processes increase the maximum energy of the particles. Thus the maximum energy is predicted to be limited by the geometry of the individual shocks, and is not expected to exceed more than a few tens of TeV. Therefore, the cores of stellar clusters are generically disfavoured as PeV CR sources. Alternative routes to the production of PeV protons include acceleration at a hypothetical large-scale wind termination shock extending beyond the core of a very compact cluster \citep{morlino2021}, or acceleration around a putative fast supernova remnant shock expanding in the turbulent magnetised environment \citep{vieu2022Emax,vieu2023}.

\section*{Acknowledgements}
This work made use of the MHD code PLUTO. Computations were performed on the HPC system Raven at the Max Planck Computing and Data Facility and on the HPC system of the Max-Planck-Institut für Kernphysik. TV and LH acknowledge J.S. Wang for helpful discussions on the implementation of the simulation. TV acknowledges S. Gabici who provided clarifications on the model by Chevalier \& Clegg (1985). 

\section*{Data Availability}
The initial conditions and output of the simulation may be shared on reasonable request to the corresponding author.



\bibliographystyle{mnras}
\bibliography{biblio} 




\appendix
\section{Analytic solution for stochastic re-acceleration}\label{app:AnalyticStochastic}
We seek a solution for Eq.~\ref{TE_stochastic_reacc} in the case where the transport coefficients are power-laws of the momentum variable, $\tau_e = \tau_0 (p/p_i)^{-\delta}$, $D_p = p^2/(2\tau_a) (p/p_i)^{-q}$.

Let us change variable from $p$ to $Z$ with $Z$ defined as:
\begin{equation}
    Z \equiv \(\frac{p}{p_i} \)^{\frac{q+\delta}{2}} \frac{1}{\eta} {\frac{2}{q+\delta}}  \, , \quad \eta = \sqrt{\tau_0/(2\tau_a)} \, ,
\end{equation}
and let us define a new function $g$ as:
\begin{equation}
    f \equiv g(Z) Z^{-\alpha} \, , \quad \alpha = \frac{3-q}{q+\delta} \, .
\end{equation}

We obtain the following equation for the function $g$:
\begin{equation}\label{eqbesselK}
    Z^2 g'' + Z g' - \( Z^2 + \alpha^2 \) g = - \tilde{Q}(p) Z^{\alpha+\frac{2q}{q+\delta}} \, ,
\end{equation}
where factors have been absorbed in $\tilde{Q}$.

The solution for a monoenergetic injection $\delta(Z-Z_0)$ is the Macdonald function (modified Bessel function of the second kind) with appropriate normalisation:
\begin{equation}
    g_H(Z) = \frac{K_{\alpha}(Z)}{K_{\alpha}(Z_0)} \frac{1}{Z_0^2+\alpha^2} \, ,
\end{equation}
and the general solution for any source function is then obtained by the following convolution:
\begin{equation} \label{finalresultf}
    f(p) \propto \int_{Z(p_i)}^{Z(p)} \frac{\dd Z_0}{Z_0^2+\alpha^2} \frac{K_{\alpha}(Z)}{K_{\alpha}(Z_0)}  \( \frac{Z}{Z_0} \)^{-\alpha} Z_0^\frac{2q}{q+\delta} \tilde{Q}\(  Z_0^{\frac{2}{q+\delta}} \) \, ,
\end{equation}
where we made the simplifying assumption that $\tilde{Q}(\lambda p) \propto \tilde{Q}(p)$ and $Q(p) =0$ if $p<p_i$. The integration can be performed numerically. The result is shown in Fig.~\ref{fig:compute_all_eta} in the case $\delta = q$, which corresponds to the case where the transport is always dominated by the interaction with magnetic modes.  We assumed that the injection is a power-law: $\tilde{Q}(P) \propto P^{-\beta}$ with $\beta =4$, as expected for instance in an environment where multiple shocks inject particles via the first order Fermi mechanism (e.g. a massive star cluster).

As discussed in the main text, the solution displays a typical bump at low energies. The position of the bump can be estimated as follows.
In the special case $q=0$, i.e. the re-acceleration and escape times do not depend on momentum, the stationary transport equation can be solved exactly and one finds a power-law solution, $f \propto P^{-\zeta}$, where:
\begin{equation}
    \zeta = 3/2 + 3/2 \( 1+ \frac{4}{9 \eta^2 }  \)^{1/2} \, .
\end{equation}
In the general case where the timescales depend on momentum, one can think of this solution applying locally, the local power-law index being:
\begin{equation}
    \zeta(p) =
    3/2 + 3/2 \( 1+ \frac{4}{9 \eta^2} \( \frac{p}{p_i} \)^{q+\delta}  \)^{1/2} \, .
\end{equation}
The position of the peak of the rescaled function $p^{\gamma} f(p)$ is determined by solving $\zeta = \gamma$, which provides:
\begin{equation}\label{bumpposition}
       p_b
      = \(
      \( \gamma - 3  \)
       \gamma \eta^2
       \)^{\frac{1}{q+\delta}} p_i \, .
\end{equation}
The theoretical position of the bump is marked by the vertical dashed lines in Figures~\ref{fig:compute_all_eta}, which show that this is indeed an excellent estimate.

At momenta smaller than $p_b$, the solution is  hard. An approximation can be obtained in this regime as follows.
In the limit $Z \ll 1$, the modified Bessel function of the second kind behaves as:
\begin{equation}
    K_\alpha(Z) \sim 2^{\alpha-1} Z^{-\alpha} \Gamma(\alpha) \, .
\end{equation}
If $\eta (q+\delta)/2 > 1$, then there exists a range of low values of $Z$ where this asymptotic expansion applies. The integrand of Eq.~\ref{finalresultf} can be simplified and the integral becomes analytic if we assume a power-law injection $\tilde{Q}\(P_0\) = P_0^{-\beta}$:
\begin{align}\label{fDplowpapprox}
    f(p) \propto 
    p^{q-3} \( \(\frac{p}{p_i}\)^{3-\beta+(q+\delta)/2} - 1 \) \, , 
\end{align}
such that for $p_i \ll p \ll \(\eta (q+\delta)/2\)^{2/(q+\delta} p_i $\, the solution is a power-law:
\begin{equation}
    f(p) \propto p^{q-3} \, .
\end{equation}
Remarkably, this asymptotic solution does not depend neither on the slope of the injection spectrum, not on the scaling of the escape time. We in fact retrieve the well-known result that stochastic re-acceleration produces a hard spectrum at low momenta. 
The approximation given by Eq.~\ref{fDplowpapprox} is shown by the dashed-dotted lines in Figures~\ref{fig:compute_all_eta}. One can check that it is indeed a good approximation up to the momentum $p = \(\eta (q+\delta)/2\)^{2/(q+\delta)} p_i $.

\bsp	
\label{lastpage}
\end{document}